\documentclass[fleqn,usenatbib]{mnras}
\usepackage{fix-cm}
\usepackage{calc}
\usepackage{newtxtext,newtxmath}

\usepackage[T1]{fontenc}

\DeclareRobustCommand{\VAN}[3]{#2}
\let\VANthebibliography\thebibliography
\def\thebibliography{\DeclareRobustCommand{\VAN}[3]{##3}\VANthebibliography}

\usepackage{graphicx}	
\usepackage{amsmath}	
\usepackage{physics}
\usepackage{mathtools}
\usepackage{color}
\usepackage[normalem]{ulem} 
\usepackage{verbatim} 

\newcommand{\bin}{{\rm b}} 
\newcommand{\fmin}{f_\mathrm{min}} 
\newcommand{\fmax}{f_\mathrm{max}} 
\newcommand{\fmid}{f_\mathrm{mid}} 
\newcommand{\Mo}{M_\odot} 
\newcommand{\alphasafety}{\alpha_\mathrm{safety}} 
\newcommand{\bilby}{\texttt{bilby}~} 
\newcommand{\secref}[1]{Sec.~\ref{#1}} 
\newcommand{\figref}[1]{Fig.~\ref{#1}} 
\newcommand{\tabref}[1]{Table~\ref{#1}} 
\newcommand{\appref}[1]{Appendix~\ref{#1}} 

\usepackage{acronym}
\newacro{GW}[GW]{Gravitational Wave}
\newacro{LVK}[LVK]{the LIGO-Virgo-KAGRA collaboration}
\newacro{BNS}[BNS]{binary neutron star}
\newacro{SVD}[SVD]{singular value decomposition}
\newacro{NPE}[NPE]{neural posterior estimation}
\newacro{JSD}[JSD]{Jensen-Shannon divergence}
\newacro{SBI}[SBI]{Simulation-based inference}
\newacro{NSF}[NSF]{Neural spline flow}
\newacro{KS}[KS]{Kolmogorov-Smirnov}
\newacro{PSD}[PSD]{power spectrum density}

\title[Simulation-based Inference: Application of Summary Data]{Simulation-based Inference for Gravitational Waves from Binary Neutron Stars: Application of Summary Data from Heterodyning}

\author[M. Iwaya et al.]
{Masaki Iwaya,$^{1}$\thanks{E-mail: IwayaM@cardiff.ac.uk (MI)}$^{,2}$
Vivien Raymond,$^{1}$
Soichiro Morisaki$^{2}$,
Kazuki Takada$^{2}$
\\
$^{1}$ Gravity Exploration Institute, School of Physics and Astronomy, Cardiff University, Cardiff, CF24 3AA, UK\\
$^{2}$ Institute for Cosmic Ray Research, University of Tokyo, Kashiwanoha 5-1-5, Kashiwa, Chiba, 277-8582, Japan
}

\date{Accepted XXX. Received YYY; in original form ZZZ}

\pubyear{\the\year{}}

\begin{document}
\label{firstpage}
\pagerange{\pageref{firstpage}--\pageref{lastpage}}
\maketitle

\begin{abstract}
Gravitational-wave parameter estimation for binary neutron star (BNS) systems poses
severe computational challenges due to the extended signal duration, which can reach several minutes in current detectors. Neural posterior estimation (NPE), a simulation-based inference approach, offers dramatic speedups but requires effective dimensionality reduction of the high-dimensional input data. We present a novel compression strategy based on likelihood-oriented summary statistics derived from the relative binning formalism of \citet{ZackayRelBin}, which compresses raw frequency-domain data into the summary data. The summary data is based on a polynomial approximation of the waveform ratio using frequency banding grounded in post-Newtonian approximation, and directly evaluated with only $O(1000)$ sample points of the waveform. As a result, both the training and storage cost become more efficient than previously reported networks for BNS inference.

We train a set of NPE networks on these summary statistics and validate a network against traditional nested sampling over 1024 BNS injections. The network produces well-calibrated posteriors across all source parameters we consider, with Jensen–Shannon divergences (JSD) consistent with numerical noise for most parameters.
Although we find that the median JSD for the most inconsistent parameter exceeds $10^{-2}$ bits with current configurations, our results show potential for rapid parameter estimation of the BNS signal.

\end{abstract}




\section{Introduction}
\label{Intro}

Since the first direct detection of \ac{GW} in 2015 \citep{GW150914} by the Laser Interferometer Gravitational-Wave Observatory (LIGO) \citep{LIGO}, \ac{LVK} \citep{Virgo, KAGRA, LVK} has observed several hundred \ac{GW} signals to date \citep{GWTC-1, GWTC-2, GWTC-2.1, GWTC-3, GWTC-4}. The direct observation of gravitational waves has expanded the horizons of various scientific fields, such as tests of general relativity \citep{TGR-1, TGR-2, TGR-3} and population inference on \ac{GW} sources \citep{RP-1, RP-2, RP-3, RP-4}.

In particular, \ac{BNS} system has beenan interesting source type because it is expected to emit not only \ac{GW} that can be observed by \ac{LVK} but also electromagnetic signals. The first detection of the set of a \ac{GW} and an associated gamma ray burst in 2017 \citep{GW170817, GW170817MM} marked the beginning of multi-messenger astronomy with \ac{GW}s, opening unprecedented opportunities to probe fundamental physics and astrophysics. Spectroscopic observations confirmed the presence of lanthanides and other heavy elements, providing the first direct evidence that \ac{BNS} is an astrophysical site for r-process nucleosynthesis \citep{GW170817:r-process}. Furthermore, the combination of the luminosity distance inferred from the \ac{GW} signal and the redshift measured from the electromagnetic counterpart enabled the first standard siren measurement of the Hubble constant \citep{GW170817:siren}. The \ac{GW} signal itself, particularly through tidal deformability measurements, provided stringent constraints on the neutron star equation of state \citep{GW170817:EoS}. Future \ac{BNS} observations are also expected to further deepen our understanding of these fields.

At the root of these scientific discoveries lies the essential need for precise measurements of the parameters of \ac{BNS}, such as masses, spins, prompt sky locations, and distances. Typically, the analysis pipeline on \ac{GW}s from compact binary coalescence relies on Bayesian inference under the assumption that the noise around the event is Gaussian and stationary \citep{LALinference, Bilby}. The posterior probability distribution of the source parameters is typically obtained through nested sampling algorithms \citep{NestedSampling}. However, \ac{BNS} signals present unique computational challenges compared to binary black hole signals. Due to their lower masses, \ac{BNS} systems remain in the detector's sensitive frequency band for much longer periods, potentially several minutes at current detector sensitivity. This extended duration increases the data volume required for analysis, leading to high computational costs for conventional methods. Furthermore, as the time period requiring computation increases, so does the probability of non-stationary noise appearing. This heightens the risk that the assumptions utilised in likelihood calculations may break down, which may affect the inference result \citep{GlitchJade2018, GlitchPayne2022, GlitchUdall2025}.

Machine learning approaches, particularly simulation-based inference (SBI) \citep{CranmerSBI}, have emerged as promising alternatives in the \ac{GW} context. Within the SBI framework, \ac{NPE} has proven particularly effective for \ac{GW} parameter estimation \citep{Green2020nov, Delaunoy2020, Green2020aug, Dax2021, Dax2023, Bhardwaj2023, Alvey2023, Dax2025, Hu2025}. \ac{NPE} trains a neural network to directly approximate the posterior distribution by learning from simulated pairs of parameters $\theta$ and corresponding data $d$. Once trained, \ac{NPE} can produce posterior samples for new observations in seconds, offering dramatic speedups compared to traditional sampling methods that require hours to days per event.

A central challenge in applying \ac{NPE} to \ac{BNS} signals is the high-dimensionality of the input data. The extended duration of \ac{BNS} signals, particularly at next-generation detector sensitivities, results in time-series or frequency-series data with too large dimensions for direct processing by neural networks. Previous \ac{BNS} studies \citep{Dax2025, Hu2025} have addressed this challenge through a two-stage data compression strategy. Both networks first apply heterodyne \citep{Heterodyne} to remove the dominant oscillatory components of the signal, followed by a substantial reduction in the dimension through multibanding \citep{Multiband1, Multiband2}. In addition, \citet{Hu2025} further employs \ac{SVD} to compress the data for next-generation detector sensitivity. These approaches compress the detector data in the frequency domain to a manageable dimension before feeding it to the neural network, which must then implicitly learn to perform matched filtering operations on the compressed representation.

In this paper, we propose an alternative dimensionality reduction strategy based on the summary-data statistic introduced by \citet{ZackayRelBin}. This statistic was originally developed as an efficient method to compute segmented matched-filter signal-to-noise ratios, based on the assumption that the ratio of two waveforms within a frequency bin can be approximated by a linear function.
We extend the original formulation to accommodate approximations using higher-order polynomials.

This approach offers several conceptual advantages for \ac{NPE}.
First, this method provides an efficient way of generating training data. As will be shown, the polynomial approximation allows the generation of summary data without the need to simulate all frequencies; instead, it is sufficient to simulate only as many points as there are in the summary data. This enables a drastic reduction in the computational cost associated with data generation and the storage of training data.
Secondly, the compression ratio achieved by this method is higher than that obtained by multibanding,
which is adopted in the previous works for BNS \citep{Dax2025, Hu2025}.
This is because, whereas multibanding assumes that the waveform is constant within a band, this method makes a more generalised assumption that the waveform can be regarded as a polynomial within a band. By allowing for variations, our compression method into summary data achieves much more compact dimensionality reduction than that of \citet{Dax2025}, which is $\sim60$ for \ac{LVK} sensitivity.
Third, the flexible dimensionality control through polynomial order allows systematic exploration of the accuracy-efficiency trade-off. Unlike multibanding, where the number of frequency bands must be tuned heuristically, the polynomial order provides simple and interpretable hyperparameters. This makes it straightforward to benchmark how the accuracy of summary data affects posterior fidelity in NPE without confounding factors. Together, these properties make the polynomial-based compression a practical choice for scaling \ac{NPE}-based \ac{GW} parameter estimation to realistic detector scenarios.

This paper is organised as follows: In \secref{sec: methods}, we review the basic theory of the \ac{SBI} and introduce the formulation of summary data statistics. In \secref{sec: Experiments}, we describe the training configuration in detail, then show the performance of the trained network. In \secref{sec: Discussion}, we summarise the results and discuss the future improvements.

\section{Methods}
\label{sec: methods}
\subsection{Simulation-Based Inference}
\label{sec: SBI}

In \ac{GW} astrophysics, the parameter estimation given \ac{GW} data is performed with a Bayesian inference:
\begin{align}
    p(\theta|d)\propto L(d|\theta)\pi(\theta),
\end{align}
where $p(\theta|d)$ is the posterior distribution of source parameters $\theta$ given the detector responce $d$, $\pi(\theta)$ is the prior distribution which embodies our knowledge on $\theta$, and $L(d|\theta)$ is the likelihood to obtain $d$ given $\theta.$
In the conventional approach, $L(d|\theta)$ is given by the assumption that noise follows a stationary Gaussian distribution. This yields $L(d|\theta)$ as follows \citep{Likelihood1, Likelihood2, Likelihood3}:
\begin{align}
    \label{eq: conventiional_likelihood}
    L(d|\theta)\propto\exp[-\frac{1}{2}\sum_f\frac{\abs{d(f)-h(f|\theta)}^2}{TS_n(f)}],
\end{align}
where $f$ denotes the frequency points, $S_n(f)$ is the one-sided \ac{PSD} of the detector noise, $T$ is the duration of data segment, and $h(f|\theta)$ denotes the modelled waveform from the \ac{GW} source with parameter $\theta$. 

\ac{SBI} is a class of statistical methods that enables parameter estimation and model comparison when the likelihood function $L(d|\theta)$ is intractable or computationally prohibitive to evaluate, but forward simulation of data from the model is feasible. By training a neural network to approximate $L(d|\theta)$, SBI methods can perform efficient Bayesian inference without requiring explicit likelihood evaluations.

This work uses \texttt{SBI} package \citep{SBIpackage} as the main body of the \ac{SBI} workflow. This package implements a suite of \ac{SBI} algorithms for \ac{NPE}, neural likelihood estimation, and neural ratio estimation. In this work, we adopt \ac{NPE}, which directly approximates the posterior distribution $p(\theta|d)$ by training a conditional density estimator $q_\phi(\theta|d)$, parameterised by neural network weights $\phi$. The training objective is to minimise the expected negative log-likelihood over simulated parameter--data pairs $\{(\theta_i,d_i)\}_{i=1}^N$ drawn from the joint distribution $\pi(\theta)L(d|\theta)$:
\begin{align}
    \mathcal{L}(\phi) = -\mathbb{E}_{\pi(\theta)L(d|\theta)}\left[\log q_\phi(\theta|d)\right],
\end{align}
which is equivalent to minimising the Kullback–Leibler divergence $D_\mathrm{KL}\qty(p(\theta|d)\,\|\,q_\phi(\theta|d))$. As the density estimator, we employ a neural spline flow \citep{NSF}, where each transformation is constructed from monotonic rational-quadratic splines. In this work, we utilise code developed in \citet{VRpastwork}, to which we have added new code for relative binning.

\subsection{Relative Binning for Simulation-Based Inference}
\label{sec: RelBin}

\subsubsection{The summary data}
In this work, we use the extended summary data formulation of \citet{ZackayRelBin}.
First, we take a frequency bin $\bin = [\fmin(\bin),\fmax(\bin))$ within which the ratio of waveforms can be approximated as a $D$-dimensional polynomial:
\begin{align}
    \label{eq: approximation by polynomial}
    \text{if $f\in \bin$,}\qq{}\text{then\ } r(f|\theta)\coloneqq \frac{h(f|\theta)}{h_0(f)} \approx \sum_{d=0}^Dr_d(\bin|\theta)\qty(f-\fmid(\bin))^d,
\end{align}
where $\fmid(\bin)$ is the median frequency of the said bin $\bin$, $\fmid(\bin)=(\fmin(\bin)+\fmax(\bin))/2$.
Note that $D$ is fixed to be $D=1$ in \citet{ZackayRelBin}. This insight was used for the fast computation of
\begin{align}
    \label{eq: (d,h)}
    \qty(d(f),h(f|\theta)) \coloneqq 4\sum_{f}\frac{d(f)h^\ast(f|\theta)}{TS_n(f)},
\end{align}
which appears when \eqref{eq: conventiional_likelihood} is expanded. Then a set of summary data
\begin{align}
    \label{eq: A_d(b)}
    A_d(\bin)&=4\sum_{f\in\bin}\frac{d(f)h^\ast_0(f)}{TS_n(f)}(f-\fmid(\bin))^d,\\
    \label{eq: B_d(b)}
    B_d(\bin)&=4\sum_{f\in\bin}\frac{\abs{h_0(f)}^2}{TS_n(f)}(f-\fmid(\bin))^d,
\end{align}
is introduced to achieve the fast evaluation of \eqref{eq: (d,h)}:
\begin{align}
    (d(f),h(f|\theta))\approx \sum_{\bin}\sum_{d=0}^Dr^\ast_d(\bin|\theta)A_d(\bin),
\end{align}
and $(h(f|\theta), h(f|\theta)):$
\begin{align}
    (h(f|\theta),h(f|\theta))\approx \sum_{\bin}\sum_{d_1=0}^D\sum_{d_2=0}^Dr^\ast_{d_1}(\bin|\theta)B_{d_1+d_2}(\bin)r_{d_2}(\bin|\theta).
\end{align}

Now we adopt these tools for our purpose: to conduct training based on summary data. This requires a set of simulation data which has the form of $\{A_d(f)\}$. In our implementation, the training dataset is generated in two stages. In the first stage, a finite number of simulated summary data from the modelled \ac{GW} waveform without noise contamination are generated, sampling several intrinsic parameters. In the second stage, the extrinsic parameters and random noise are added during each training iteration. This generates a dataset possessing semi-infinite diversity from a finite number of intrinsic parameter samples.

In the first stage, chirp mass $\mathcal{M}=(m_1m_2)^{3/5}/(m_1+m_2)^{1/5}$, mass ratio $q=m_2/m_1\leq1$, spin parameters $\vec{a}_1,\vec{a}_2$, tidal parameters $\Lambda_1,\Lambda_2$, total angular momentum azimuth $\phi_{JL}$, and phase $\phi$ are generated from their corresponding prior distributions, and summary data for these samples is created. We will denote them as $\theta_1=\{\mathcal{M},q,\vec{a}_1,\vec{a}_2,\Lambda_1,\Lambda_2,\phi_{JL},\phi\}$ for the rest of this paper.

For each sampled parameter $\theta_1$, we evaluate $h(f|\theta_1)$. Note that, however, given the assumption that the behaviour of $r(f|\theta_1)$ can be approximated as a polynomial for frequency bins, we no longer need to sample over the entire frequency domain. Instead, we only have to $D+1$ points per frequency bin to determine the approximant coefficients $\{r_0(\bin|\theta_1)\ldots r_D(\bin|\theta_1)\}$. This reduces the total points to be computed from $O(TF_\mathrm{max})$ to $O(DN_\mathrm{bin}),$ where $F_\mathrm{max}$ is the largest frequency in the strain and $N_\mathrm{bin}$ is the number of frequency bins. We will further discuss the efficiency of this reduction in \secref{sec: compression}.
Practically, we evaluate $\{r_d(\bin)\}$ by solving
\begin{align}
    \qty(
    \begin{array}{c}
        r(\fmin(\bin)|\theta_1)\\\vdots\\ r(\fmax(\bin)|\theta_1)
    \end{array}
    )=\mathbf{V}(\bin)
    \qty(
    \begin{array}{c}
        r_0(\bin|\theta_1)\\\vdots\\r_D(\bin|\theta_1)
    \end{array})
    ,
\end{align}
where $\mathbf{V}(\bin)$ is the $(D+1)\times(D+1)$-shaped Vandermonde matrix of $\bin$:
\begin{align}
    \qty[\mathbf{V}(\bin)]_{i,j} = \qty(\fmin(\bin)+\frac{i}{D}\qty(\fmax(\bin)-\fmin(\bin)))
^{j},
\end{align}
where $i,j$ both runs $0,1,\ldots,D.$ Then the zero-noise summary data from $h(f|\theta_1)$ can be computed as
\begin{align}
    H_d(\bin|\theta_1) &\coloneqq 4\sum_{f\in\bin}\frac{h(f|\theta_1)h_0^\ast(f)}{TS_n(f)}(f-\fmid(\bin))^d\label{eq: definition of H(b)}\\
    &\approx 4\sum_{f\in\bin}\sum_{k=0}^D\frac{\qty[r_d(\bin|\theta_1)h_0(f)]h_0^\ast(f)}{TS_n(f)}(f-\fmid(\bin))^{d+k}\notag\\
    &=\sum_{k=0}^Dr_d(\bin|\theta_1)B_{d+k}(\bin),
\end{align}
where we use \eqref{eq: approximation by polynomial} and \eqref{eq: B_d(b)}. Therefore, we can express the construction of summary data as:
\begin{align}
    \label{eq: How H(b) is constructed}
    \mathbf{H}(\bin|\theta_1) \coloneqq \qty(
    \begin{array}{c}
        H_0(\bin|\theta_1)\\\vdots\\H_D(\bin|\theta_1)
    \end{array}
    ) = \mathbf{B}(\bin)\qty(\mathbf{V}(\bin))^{-1}\qty(
    \begin{array}{c}
        r(\fmin(\bin)|\theta_1)\\\vdots\\ r(\fmax(\bin)|\theta_1)
    \end{array}
    ),
\end{align}
where $\mathbf{B}(\bin)$ is pre-computable $(D+1)\times(D+1)$-shaped matrix which satisfies
\begin{align}
    \qty[\mathbf{B}(\bin)]_{i,j} = B_{i+j}(\bin)=4\sum_{f\in\bin}\frac{\abs{h_0(f)}^2}{TS_n(f)}(f-\fmid(\bin))^{i+j}.
\end{align}
We store the summary data generated in this manner as the first stage of generation, to be used later for considering the effects of further parameters.

Parameters not used in generating the training data at the first stage are dynamically generated during the training process and applied to the training data. This enables the preparation of a semi-infinite number of training datasets. The parameters sampled here comprise right ascension $\alpha$, declination $\delta$, polarisation $\psi$, luminosity distance $d_L$, and time shift from the trigger time $\Delta t$. We will denote them as $\theta_2$. The application of the effect of $\theta_2$ is simple except for time shifting; time shifting produces different effects at each frequency, making it complex in the summary data. As long as the time difference is so small that the polynomial approximation holds both before and after the time shift of $\Delta t$, we have
\begin{align}
    \label{eq: time shift on r(f|theta)}
    \qty(
    \begin{array}{c}
        r(\fmin(\bin))\exp[-2\pi i \fmin(\bin)\Delta t]\\\vdots\\ r(\fmax(\bin))\exp[-2\pi i\fmax(\bin)\Delta t]
    \end{array}
    )=\mathbf{V}(\bin)
    \qty(
    \begin{array}{c}
        r_0\qty(\bin|\theta_1+\{\Delta t\})\\\vdots\\r_D\qty(\bin|\theta_1+\{\Delta t\})
    \end{array}).
\end{align}

By combining \eqref{eq: How H(b) is constructed} and \eqref{eq: time shift on r(f|theta)}, one finds that a time shift can be performed by a linear transformation:
\begin{align}
    \mathbf{H}(\bin|\theta+\{\Delta t\})
    =\mathbf{B}(\bin)\qty(\mathbf{V}(\bin))^{-1}\mathbf{T}(\Delta t)\mathbf{V}(\bin)\qty(\mathbf{B}(\bin))^{-1}\mathbf{H}(\bin|\theta),
\end{align}
where we define
\begin{align}
    &\mathbf{H}(\bin|\theta+\{\Delta t\}) \coloneqq
    \qty(
    \begin{array}{c}
        H_0(\bin|\theta_1+\{\Delta t\})\\\vdots\\H_D(\bin|\theta_1+\{\Delta t\})
    \end{array}
    ),\\
    &\mathbf{T}(\Delta t)\coloneqq \operatorname{diag}(\exp(-2\pi i\fmin(\bin)\Delta t),\ldots,\exp(-2\pi i\fmax(\bin)\Delta t)).
\end{align}

By the methodology described thus far, one can prepare a semi-infinite number of zero-noise summary data sets. From here, we explain how to introduce noise into these data sets. The simplest method for generating simulated noise is to add idealised noise at each frequency and then apply compression in the same manner as when constructing a summary dataset. However, this method can be very inefficient considering the efficiency of the compression. Therefore, we introduce an efficient technique applicable when we assume that the noise is stationary Gaussian. If we assume so, noise $n(f)$ has the following properties:
\begin{align}
    &\langle n(f) \rangle = 0,\qq{} \langle n(f)n(f')\rangle = \langle n^\ast(f)n^\ast(f')\rangle = 0,\\ &\langle n(f)n^\ast(f') \rangle = \frac{1}{2}TS_n(f)\delta(f-f'),
\end{align}
for all $f,f'>0$. Here, $\langle\cdot\rangle$ means the ensemble mean. By exploiting these properties, for summary data consisting solely of noise, 
\begin{align}
    N_d(\bin) \coloneqq 4\sum_{f\in\bin}\frac{n(f)h_0^\ast(f)}{TS_n(f)}(f-\fmid(\bin))^d,
\end{align}
we have
\begin{align}
    &\langle N_d(\bin) \rangle = 0,\qq{}\langle N_{d_1}(\bin)N_{d_2}(\bin')\rangle =\langle N^\ast_{d_1}(\bin)N^\ast_{d_2}(\bin')\rangle =0,\\
    &\langle N_{d_1}(\bin)N^\ast_{d_2}(\bin')\rangle =2 B_{d_1+d_2}(\bin)\delta_{\bin,\bin'},
\end{align}
where
\begin{align}
    \delta_{\bin,\bin'}=\begin{cases}
        1&\text{if\ } \bin = \bin',\\
        0&\text{otherwise}.
    \end{cases}
\end{align}
Therefore, the noise summary data can be simulated using a multivariate normal distribution with a mean of zero vector and a covariance matrix of $\mathbf{B}(\bin)$ for each frequency bin. Furthermore, if we take $\mathbf{W}(\bin)$ which satisfies $\mathbf{W}(\bin)\qty(\mathbf{W}(\bin))^{T}=\qty(\mathbf{B}(\bin))^{-1}$, the whitening transformation,
\begin{align}
    \mathbf{H}(\bin|\theta) \mapsto \mathbf{W}(\bin)\mathbf{H}(\bin|\theta),
\end{align}
makes the addition of noise summary data simply adding a standard uncorrelated Gaussian noise for every $\bin$ and dimensionality.

Finally, for the inference on actual data, we have to project the frequency-domain strain data onto the summary data space. However, this is a straightforward step, as this can be achieved by simply computing the summary data through the definition \eqref{eq: A_d(b)}, then applying the whitening transformation.

\subsubsection{The determination of the frequency bins}

To accomplish the transformations, we need to split the entire frequency domain into frequency bins. In this work, we determine the frequency bin sequentially based on the phase evolution of the quadrupole \ac{GW} mode:
\begin{align}
    h(f|\theta)&=A(f,\theta)\exp[-i\Psi(f|\theta)]\notag\\&=A(f,\theta)\exp[-i\sum_i\psi_i(\theta)f^{\gamma_i}],
\end{align}
where $A(f,\theta)$ is a real function that determines the amplitude, and $\Psi(f|\theta)$ is another real function that describes the phase evolution of the \ac{GW}. Specifically, we approximate $\Psi(f|\theta)$ as follows,
\begin{align}
    &\Psi(f|\theta)\notag\\
    \approx&\psi_0(\theta)f^{-\frac{5}{3}}+\psi_2(\theta)f^{-1}+\psi_3(\theta)f^{-\frac{2}{3}}+\psi_5(\theta)f+\psi_{10}(\theta)f^{\frac{5}{3}},
\end{align}
where, in $c=G=1$ unit system,
\begin{align}
    &\psi_0(\theta)=\frac{3}{128}(\pi\mathcal{M})^{-\frac{5}{3}}\\
    &\psi_2(\theta)=\frac{5}{96}\qty(\frac{743}{336}+\frac{11}{4}\eta)\eta^{-\frac{2}{5}}(\pi\mathcal{M})^{-1}\\
    &\psi_3(\theta)=\frac{3}{32}\qty(\beta-4\pi)\eta^{-\frac{3}{5}}\qty(\pi\mathcal{M})^{-\frac{2}{3}}\\
    &\psi_5(\theta) = 2\pi t_\mathrm{c}\\
    &\psi_{10}(\theta)=-\frac{117}{256}\eta^{-2}\tilde{\Lambda}(\pi\mathcal{M})^{\frac{5}{3}},
\end{align}
and
\begin{align}
    &\eta=\frac{q}{(1+q)^2}\\
    &\beta=\frac{1}{12(1+q)^2}\qty[\qty(\frac{113}{12}+\frac{25q}{4})\chi_{1z}+\qty(\frac{113}{12}+\frac{25}{4q})q^2\chi_{2z}]\\
    &\tilde{\Lambda}=\frac{16}{13}\frac{(1+12q)\Lambda_1+(q+12)q^4\Lambda_2}{(1+q)^5},
\end{align}
and $\chi_{iz}$ is the $z$-component of the individual spin, $\vec{a_i}$. We impose the following condition on the bins:
\begin{align}
    \max_\theta\abs{\Delta\Psi(\fmax,\theta)-\Delta\Psi(\fmin,\theta)}\ll1,
\end{align}
where we define $\Delta\Psi(f|\theta)$ as the phase of waveform ratio:
\begin{align}
    \Delta\Psi(f|\theta) = \Psi(f|\theta)-\frac{3}{128}(\pi\mathcal{M}_0f)^{-\frac{5}{3}}-2\pi t_\mathrm{c,0}f,
\end{align}
where $\mathcal{M}_0$ and $t_\mathrm{c,0}$ are the reference values with which we evaluate the waveform ratio. We provide them as the median values of the maximum and minimum values of the prior range.

For each frequency bin $\bin$, we set the $\fmin(\bin)$ as the smallest frequency point among the frequency points that are not a member of frequency bin yet. We achieve the condition by splitting the frequency range into frequency bins by giving $\fmax$ as the maximum frequency that satisfies
\begin{align}
    \label{eq: introduction of alpha safety}
    \alphasafety\cdot\qty(\fmax -\fmin)<\frac{2\pi}{\max_{i,\theta}\abs{\delta\psi_i(\theta)\gamma_i\fmin^{\gamma_i-1}}},
\end{align}
where $\delta\psi_i(\theta)$ is the coefficient of $\Delta\Psi(f|\theta)$ as the series of $f$, and $\alphasafety$ is a tunable parameter to further ensure that the polynomial approximation holds. This criterion is easy to compute but can be stricter than that proposed in \citet{ZackayRelBin}. For each $i$, the $\theta$ that gives the maximum $\delta\psi_i(\theta)$ can be analytically derived from the given prior of $\theta$.

\subsection{Compression of summary data in practice}
\label{sec: compression}
\begin{figure}
	\includegraphics[width=\columnwidth]{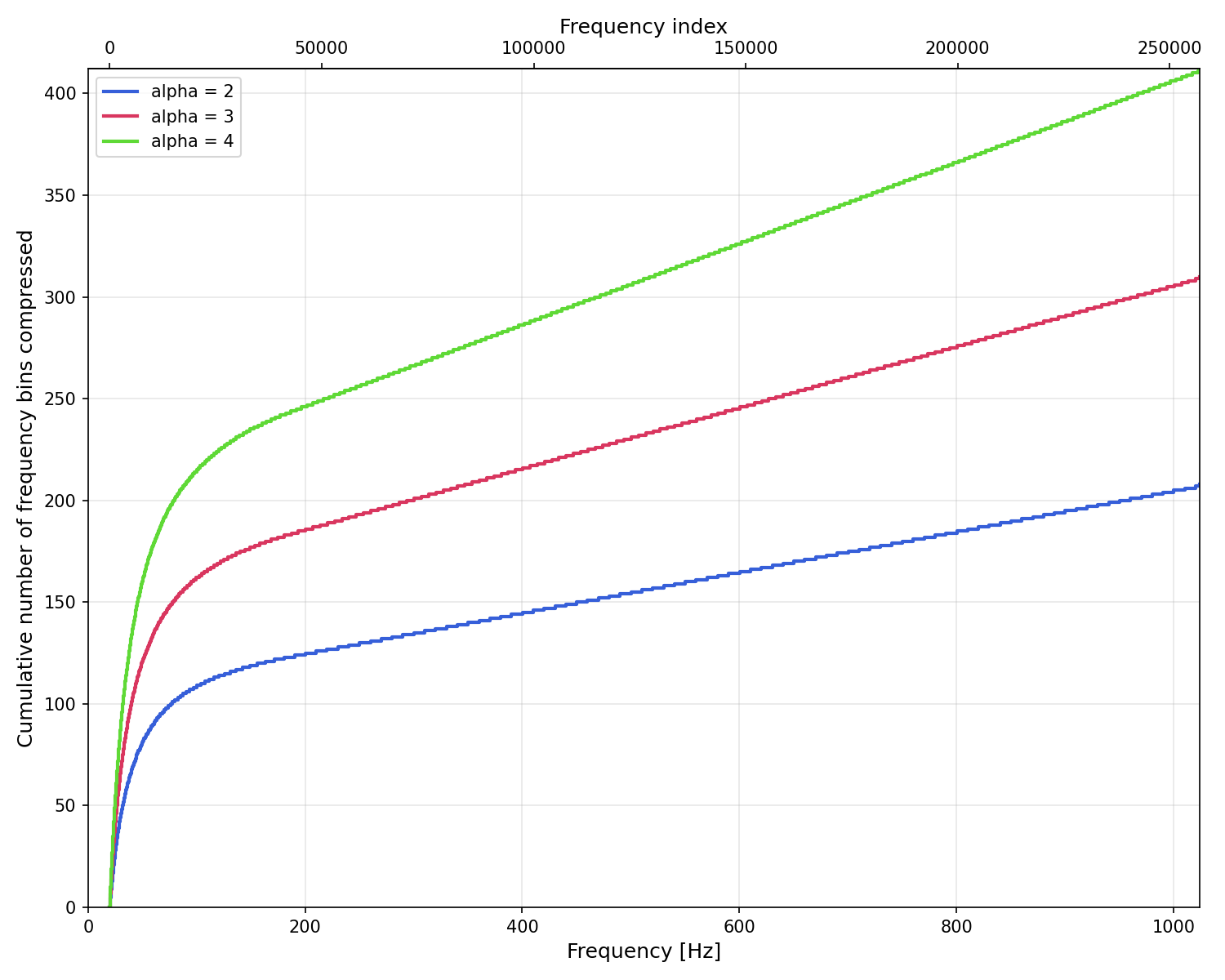}
    \caption{Compressions of the frequency domain by the relative binning scheme implemented in this work. The horizontal axes show the frequency (lower axis) or corresponding frequency index (upper axis), while the vertical axis indicates the frequency bin to which the frequency index belongs.}
    \label{fig:binning scheme}
\end{figure}

To test the validity of the summary data statistics, we investigate how data is actually compressed using our method for $\alphasafety =2,3,4$ cases. As shown in \figref{fig:binning scheme}, our algorithm assigns 257025 frequency indices, which represent the frequency range of $[20,1024]$ Hz with the resolution of $1/256$ Hz, to $[208, 310, 412]$ bins respectively for $\alphasafety =2,3,4$. If we define the compression factor by the ratio between the dimensionalities before/after the compression, the compression factor from our treatment exceeds at least 200. This is at least three times as efficient as the multibanding adopted in \citet{Dax2025}. \tabref{tab: compression summary} summarises the compression factor achieved in our configurations.

As expected, the stricter the binning threshold is, the greater the number of frequency bins. The figure illustrates this assignment by plotting the bin number (vertical axis) against the frequency index (lower horizontal axis) or physical frequency (upper horizontal axis). As the figure shows, the frequency bin resolution is finer at lower frequencies and becomes exponentially coarser towards higher frequencies. In particular, we observe a linear relationship between the number of bins and frequency is observed for frequencies beyond $\sim150$ Hz, under the prior range and $\alphasafety$ settings examined here. This stems from the uncertainty in $t_\mathrm{c}$, as this becomes the most contributing factor to the oscillation of frequency domain waveform, $\exp(2\pi if\Delta t_\mathrm{c})$, for sufficiently large frequency $f$.

\begin{figure}
	\includegraphics[width=\columnwidth]{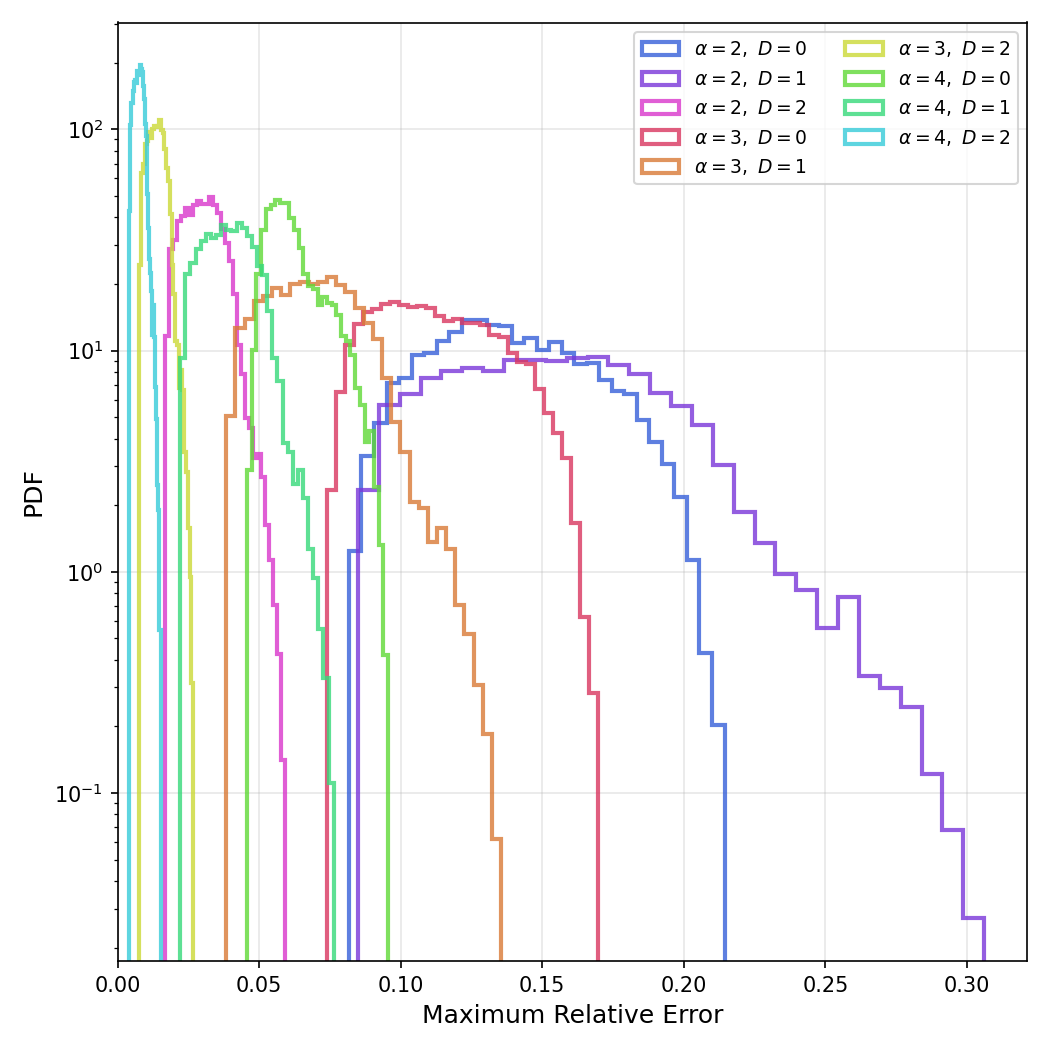}
    \caption{
    Distribution of relative errors of summary data $\{\varepsilon(\theta_{1, i})\}$ between the methods computed directly from the definition or polynomial approximation using fewer points.
    }
    \label{fig: relative errors}
\end{figure}

To validate the polynomial approximation, we generate summary data for zero-noise waveforms across 20,000 intrinsic parameter sets $\{\theta_{1, i}\}$. For each parameter set $\theta_{1, i}$, the summary data consists of $O(100)$ to $O(1000)$ complex values, compressed from the original waveform frequency series of $O(10^5)$ points. We compare the summary data $\{H_d(\bin|\theta_{1,i})\}$ computed via the polynomial approximation \eqref{eq: How H(b) is constructed}, $\{H_d(\bin|\theta_{1,i})\}_\mathrm{approx}$, against that obtained from the exact definition \eqref{eq: definition of H(b)}, $\{H_d(\bin|\theta_{1,i})\}_\mathrm{exact}$. As a figure of merit, for each sampled $\theta_{1,i}$, we compute the maximum relative error over all components,
\begin{align}
    \label{eq: Definition of varepsilon}
    \varepsilon(\theta_{1,i})\coloneqq\max_{d,\bin}\frac{\abs{H_d(\bin|\theta_{1,i})_\mathrm{exact}-H_d(\bin|\theta_{1,i})_\mathrm{approx}}}{\abs{H_d(\bin|\theta_{1,i})_\mathrm{exact}}},
\end{align}
and examine its distribution across the sampled parameter sets $\{\theta_{1,i}\}.$

\figref{fig: relative errors} shows the distributions of $\varepsilon(\theta_{1,i})$ across various configs for $\alphasafety,D.$ The average values are summarised in \tabref{tab: compression summary}.
As expected, a stricter threshold (or larger $\alphasafety$) generally yields smaller relative errors. A notable exception arises at $\alpha=2$, where the linear approximation ($D=1$) produces larger relative errors than the point approximation (corresponding to $D=0$). This anomaly suggests that introducing a linear approximation under a non-strict threshold may deteriorate accuracy, potentially due to the inadequate treatment of the behaviour of the waveform ratio in the complex plane.
We found that changing $D$ significantly affects the likelihood accuracy. Within the scope of the present experimental setup, increasing $D$ from $D=1$ to $D=2$ reduces the mean relative error to approximately 20\% of its value at $D=1$. Combining the effects of both parameters, the transition from the coarsest to the finest approximation configuration increases the input dimensionality by a factor of approximately 8, while reducing the error by a factor of approximately 20.

\begin{table}
	\centering
	\caption{Summary of the compression result we find. $\{\varepsilon(\theta_{1,i})\}$ is defined as \eqref{eq: Definition of varepsilon}. 
    }
	\label{tab: compression summary}
	\begin{tabular}{cccc}
		\hline
        $\alphasafety$& $D$ &Compression factor & observed average of $\{\varepsilon(\theta_{1,i})\}$\\
		\hline
        2&0&1236&$1.395\times10^{-1}$\\
        2&1&618&$1.555\times10^{-1}$\\
        2&2&412&$3.052\times10^{-2}$\\
        3&0&829&$1.138\times10^{-1}$\\
        3&1&414&$6.991\times10^{-2}$\\
        3&2&276&$1.387\times10^{-2}$\\
        4&0&624&$6.306\times10^{-2}$\\
        4&1&312&$3.995\times10^{-2}$\\
        4&2&208&$7.667\times10^{-3}$\\
		\hline
	\end{tabular}
\end{table}

\section{Network performances}
\label{sec: Experiments}
We trained an \ac{NPE} with a summary data scheme to explore the possibility of a network with the input of summary data statistics.

\subsection{Training configuration}
\label{sec: config}
Our \ac{NPE} is trained with simulated BNS waveforms generated with the \texttt{TaylorF2} approximant available in the \texttt{Bilby\_pipe} package \citep{bilby_pipe_paper}. Following the method described in \secref{sec: RelBin}, we generate the simulation dataset for training in two stages, adopting different extrinsic parameters for each epoch. For simplicity, we fix six of the 17 BNS parameters to zero in this study: the $x$- and $y$-components of both spins and both tidal deformabilities. Furthermore, we assume that the waveform has been observed only at LIGO Hanford. We sample the waveforms only at a limited set of frequency points and generate summary statistics from these samples. After whitening the summary data, we add uncorrelated Gaussian noise. This procedure enables efficient generation of training data without having to produce the complete time series for the entire BNS dataset. We provide the summary-data related parameters as $\alpha=2, D=0$. we generate simulation data consisting of $2^{24} \approx 1.6\times10^7$ samples and performed training.

The prior distributions for each parameter broadly follow those adopted in the \ac{LVK} works. Specifically, the masses $m_1$ and $m_2$ are uniform in the mass plane, but their boundaries are constrained by the chirp mass and mass ratio as $\mathcal{M}\in[2, 2.01] \Mo$ and $q\in[0.3,1]$, respectively, while the extrinsic parameters are set to be uninformative. This narrow chirp mass range is designed to be used in conjunction with an established approach of \citet{Dax2025}, in which only chirp masses are roughly estimated to identify a reference mass.
Regarding the prior distribution for distance, adopting the conventional uniform-in-comoving-volume-and-time distribution risked producing fewer samples at close distances. Therefore, we instead employ a uniform distribution in this study. The summary of prior distribution adopted in this study is shown in \tabref{tab: prior configuration}.

\begin{table}
	\centering
	\caption{Prior distribution for \ac{GW} source parameters we adopt in this work. Here, $U(a,b)$ denotes the one-dimensional uniform distribution over the interval $(a,b)$, $\mathrm{Sine}(a,b)$ denotes a probability density that satisfies $p(x)\propto \sin(x)$ for $x\in(a,b)$ and $p(x)=0$ otherwise, and $\mathrm{Cosine}(a,b)$ denotes a probability density that satisfies $p(x)\propto \cos(x)$ for $x\in(a,b)$ and $p(x)=0$ otherwise. See the text for the mass prior.}
	\label{tab: prior configuration}
	\begin{tabular}{cc}
		\hline
        Parameter & Prior distribution\\
		\hline
        $\mathcal{M}$ & See the text\\
        $q$ & See the text\\
        $\chi_{1x}$ & $\delta(\chi_{1x})$\\
        $\chi_{1y}$ & $\delta(\chi_{1y})$\\
        $\chi_{1z}$ & 
        $U(-0.05,0.05)$\\
        $\chi_{2x}$ & $\delta(\chi_{2x})$\\
        $\chi_{2y}$ & $\delta(\chi_{2y})$\\
        $\chi_{2z}$ & 
        $U(-0.05,0.05)$\\
        $\Lambda_1$&$\delta(\Lambda_1)$\\
        $\Lambda_2$&$\delta(\Lambda_2)$\\
        $\phi_{JL}$&$\mathrm{Sine}(0,\pi)$\\
        $\phi$&$U(0,2\pi)$\\
        $t_\mathrm{c}$&$U(t_\mathrm{c,0}-0.01 \mathrm{\ s},t_\mathrm{c,0}+0.01 \mathrm{\ s})$\\
        $d_L$&$U(50,150)$\\
        $\alpha$&$U(0,2\pi)$\\
        $\delta$&$\mathrm{Cosine}(-\frac{\pi}{2},\frac{\pi}{2})$\\
        $\psi$&$U(0,2\pi)$\\
		\hline
	\end{tabular}
\end{table}

\subsection{Validation}
We validate our networks by comparing the traditional nested-sampling algorithm method implemented in \bilby pipeline \citep{Bilby, Dynesty}. We generate 1024 BNS injection sets with additive Gaussian noise. For each injection, we generate neural posterior samples using our trained networks. We also perform the \bilby run for the same injection sets to directly compare the set of posterior samples.
In the \bilby run, we make use of multibaded likelihood evaluation \citep{Multiband2} to accelerate the evaluation. 

\begin{figure}
	\includegraphics[width=\columnwidth]{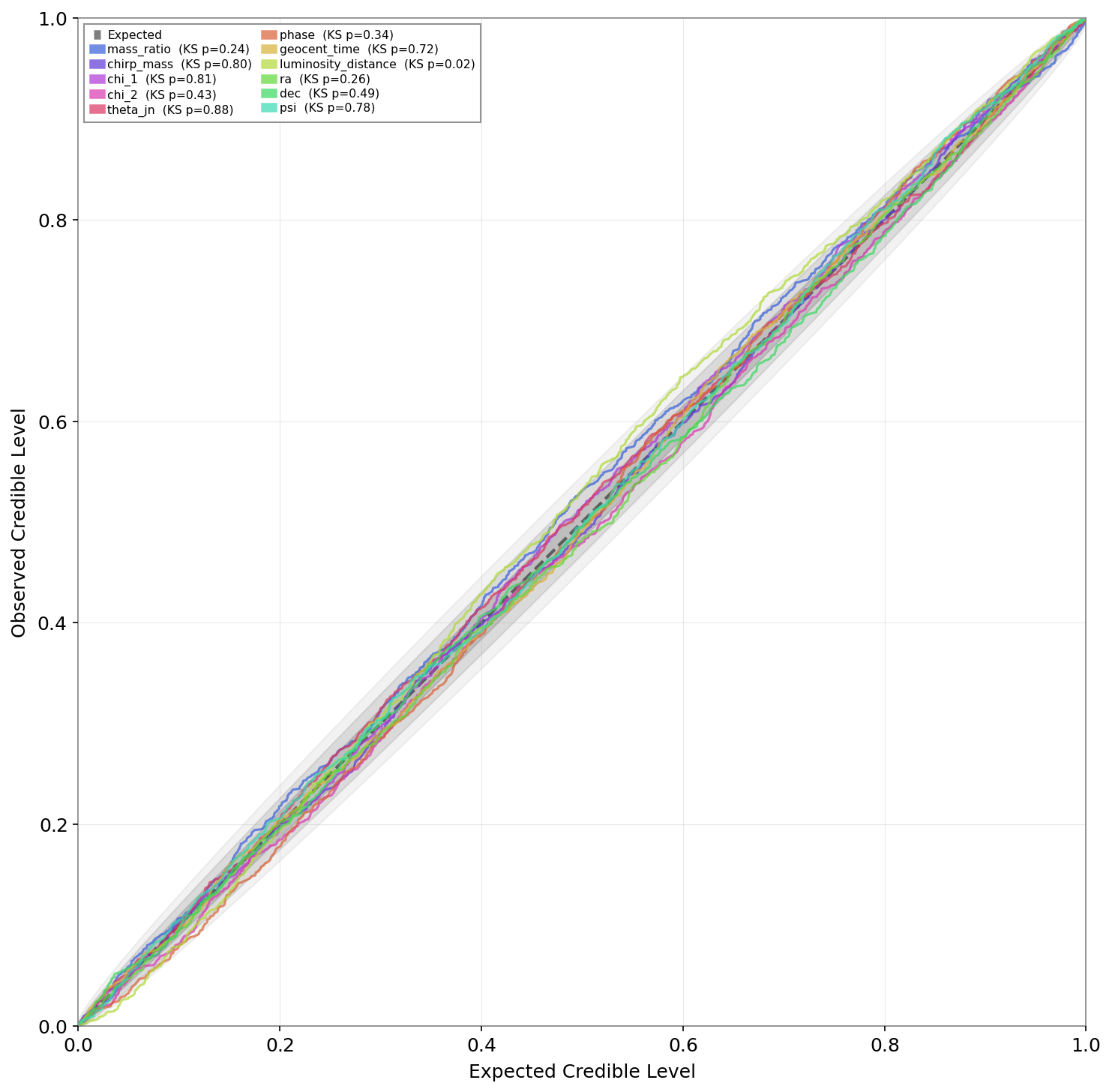}
    \caption{Percentile-percentile (PP) plots obtained from 1024 BNS injection sets. posteriors are evaluated with $\alpha=2,D=0$ configuration. The shaded areas indicate the $1$, $2$ and $3$-sigma deviations predicted by the Clopper–Pearson intervals.}
    \label{fig: pp_plot}
\end{figure}

First, we generate a percentile-percentile comparison with the injection sets. \figref{fig: pp_plot} shows the percentile-percentile plot we obtain from 1024 BNS injections. We conduct the \ac{KS} test on each parameter to evaluate whether the observed credible levels follow the expected uniform distribution under a well-calibrated pipeline. All parameters yielded \ac{KS} p-values ranging from 0.02 to 0.92. The credible levels are broadly consistent with the expected uniform distribution, with all curves lying within the $3\sigma$ Clopper–Pearson intervals.
The relatively low p-value of 0.02 warrants careful interpretation. When testing $N=11$ parameters simultaneously, the probability of obtaining at least one p-value below $\alpha=0.05
$ purely by chance is
$1 - 0.95^{11} \approx 0.43$,
meaning there is a $\sim 43\%$ chance of a false positive. The observed p-value of 0.02 does not constitute evidence of miscalibration once this multiple-comparison effect is accounted for.

\begin{figure}
	\includegraphics[width=\columnwidth]{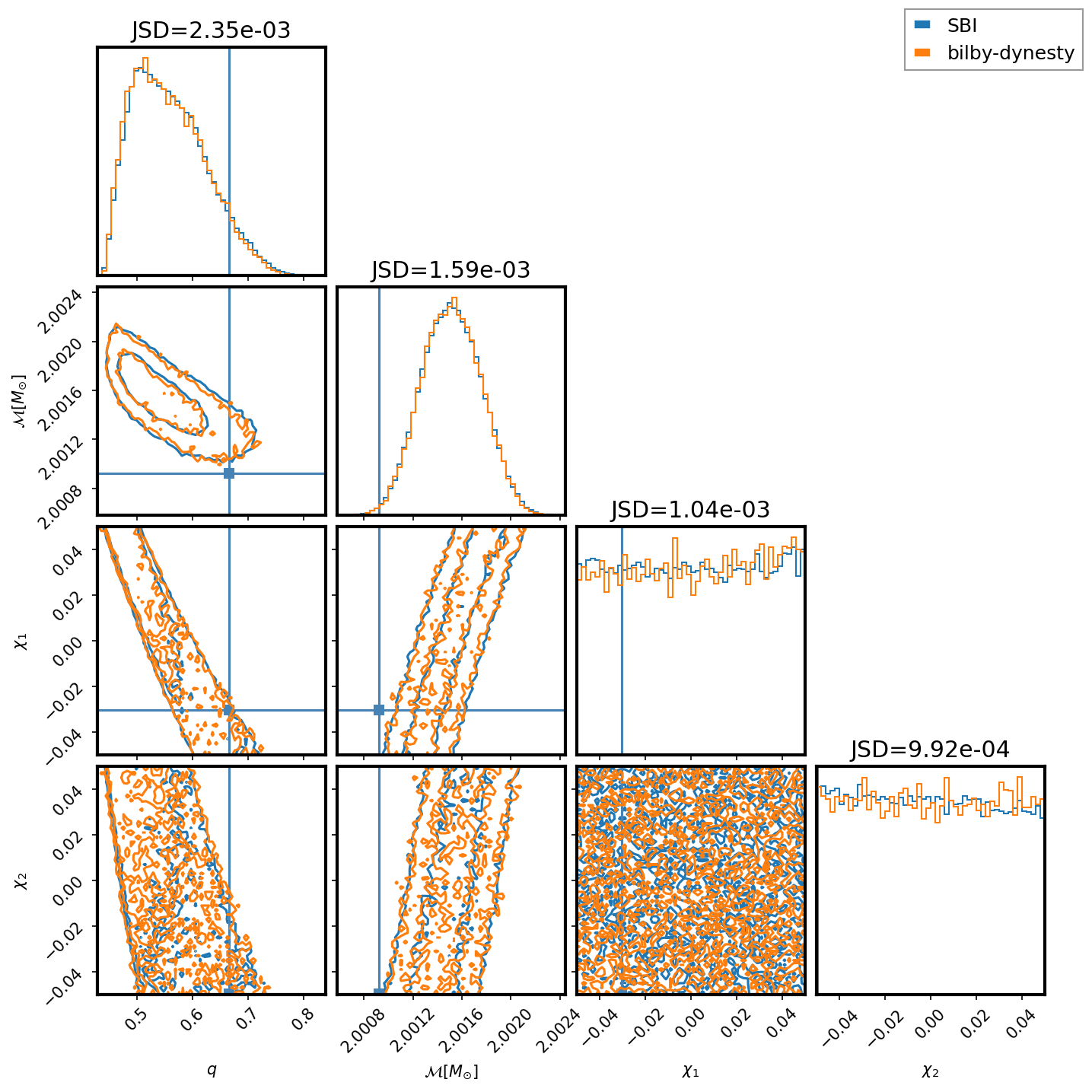}
    \caption{Corner plot for a BNS injection. For this injection, the JSD between our posterior distribution and the conventional method's chirp mass distribution exhibited one of the lowest values.} 
    \label{fig: Best_chirp_mass}
\end{figure}
\figref{fig: Best_chirp_mass} shows the corner plot for an injection that marked one of the lowest JSD for chirp mass distributions overall. The figure exhibits the strong agreement between the neural posterior and the traditional posterior.

\begin{figure}
	\includegraphics[width=\columnwidth]{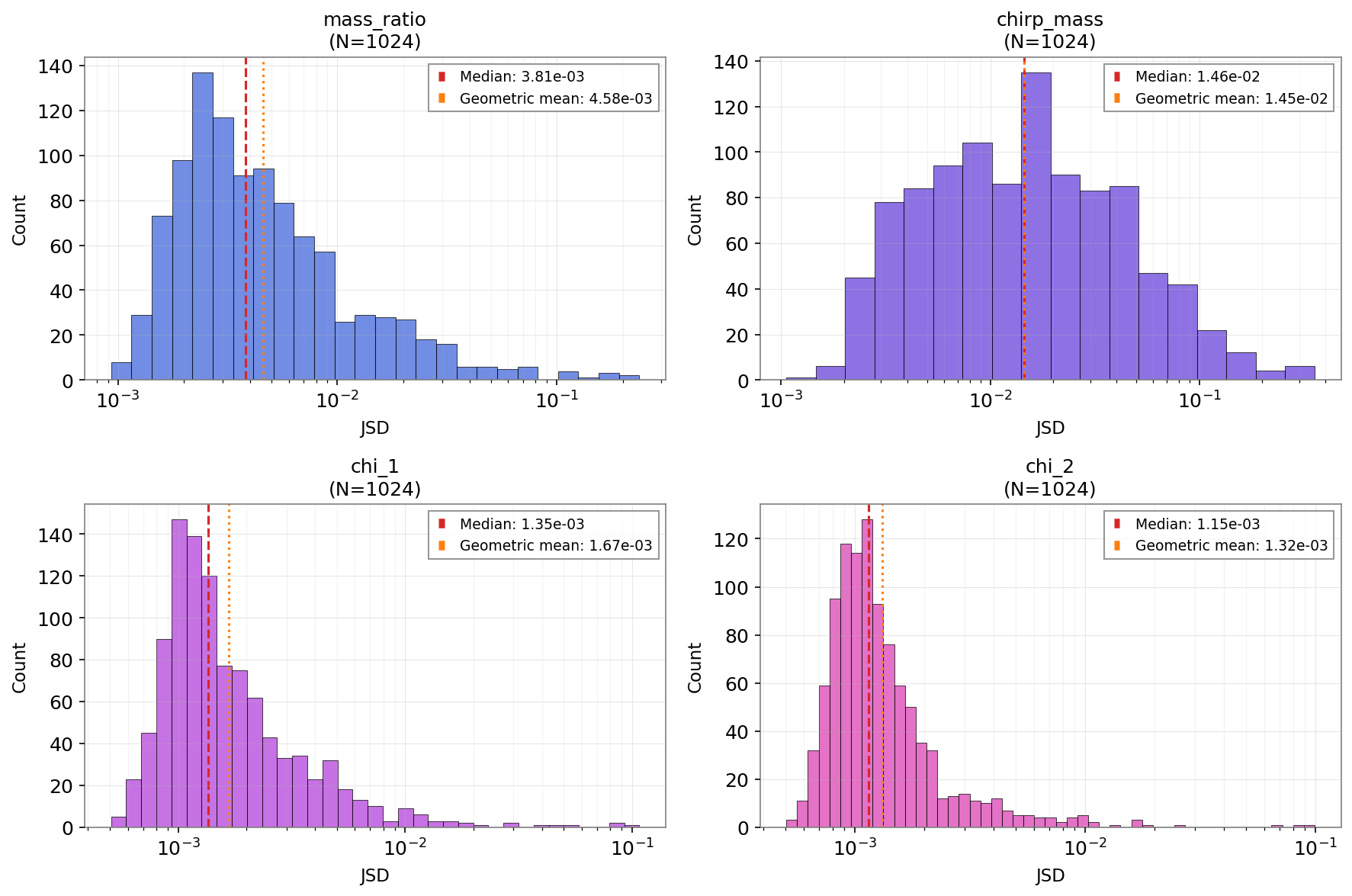}
    \caption{\ac{JSD} histogram comparing our posterior samples and traditional nested sample against 1024 injection datasets.}
    \label{fig: JSD_histogram}
\end{figure}
\figref{fig: JSD_histogram} shows the distribution of the Jensen-Shannon divergence (\ac{JSD}) for the intrinsic parameters. \ac{JSD} roughly indicates how consistent the two distributions are. Throughout this paper, we use \ac{JSD} with units of bits. In other words, the base-2 logarithm is used when evaluating \ac{JSD}. With the $\alpha=2, D=0$ network, we achieve the median \ac{JSD} of $O(10^{-3})$ for mass ratio and spin parameters. For the mass ratio and spin parameters, we achieved a median JSD of $O(10^{-3})$. Given the number of samples used to calculate these values, namely $\sim 10000$ for \bilby samples and $10^5$ samples for neural posterior, a JSD of this magnitude implies that the difference between the two posterior distributions is limited to the level of numerical error. Based on \ac{JSD} values, the most inconsistent parameter is chirp mass, which does not achieve the median \ac{JSD} below $10^{-2}$. 
\begin{figure}
	\includegraphics[width=\columnwidth]{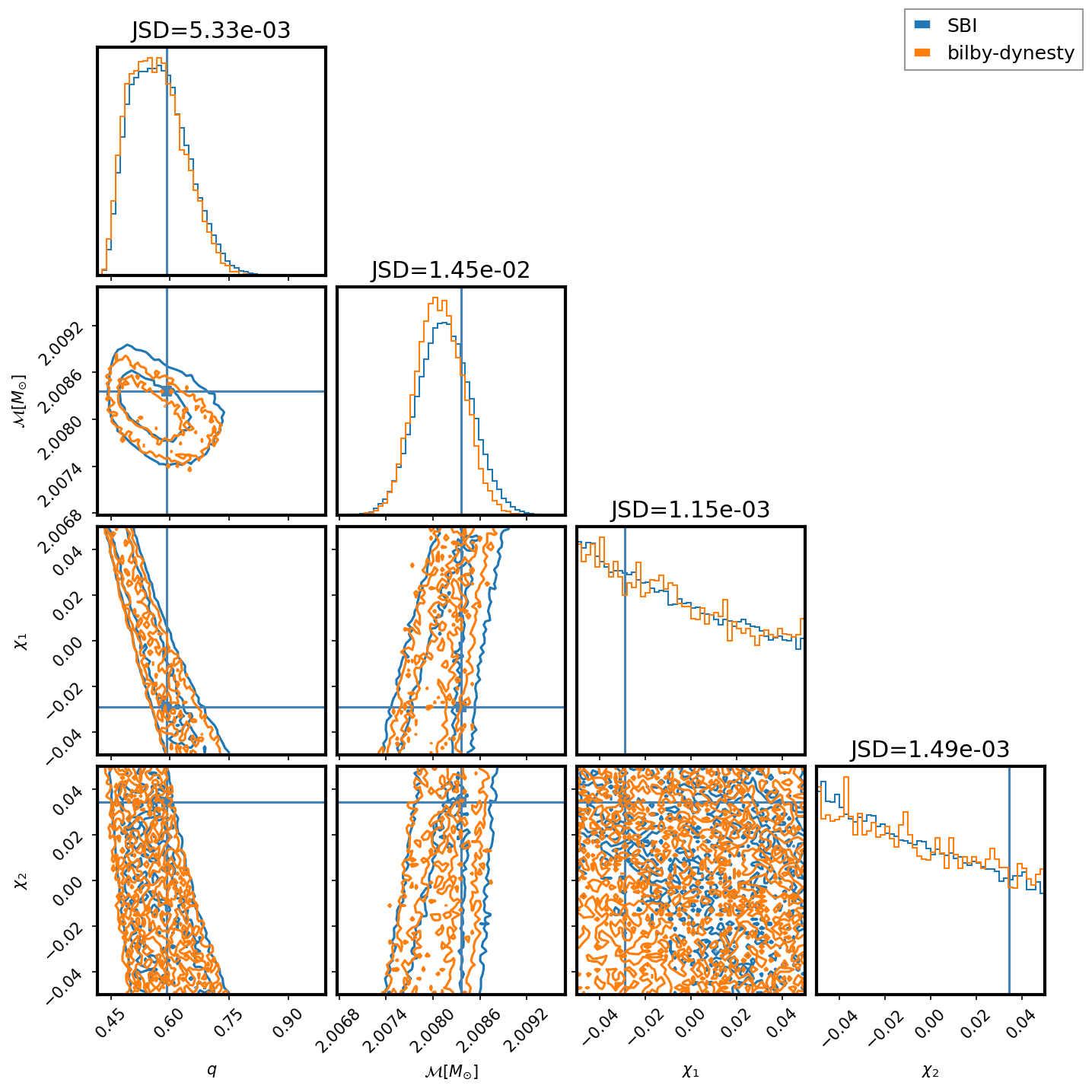}
    \caption{Corner plot for another BNS injection. For this injection, the JSD between our \ac{NPE} and the conventional method mark the median JSD obtained in this study.} 
    \label{fig: Med_chirp_mass}
\end{figure}
\figref{fig: Med_chirp_mass} shows the results of the inference where the chirp mass JSD is close to the median. For injections where the JSD is around $10^{-2}$, the NPE results show a distribution where the weights are concentrated at points slightly offset from the \bilby results for both chirp mass and mass ratio, as illustrated in this plot. Consequently, one possible approach to improving this would be to further modify the range of chirp mass. If this were applied, it would likely involve constructing a network that employs multiple reference chirp masses for a composite estimation. For the JSD histogram for every 11 parameters, see \appref{sec: Appendix}.

\section{Discussion}
\label{sec: Discussion}

In summary, we have demonstrated that \ac{NPE} trained on summary statistics achieves calibration and accuracy comparable to traditional nested sampling for \ac{BNS} parameter estimation. The PP test over 1024 injections confirms that the posteriors are well-calibrated, and the \ac{JSD} analysis shows agreement at the level of numerical noise for most parameters. The chirp mass posterior remains the primary challenge; however, the observed overlap pattern indicates that importance sampling can be applied with high efficiency in this regime. We wil further discuss the effectiveness of importance sampling in \secref{sec: Appendix_IS}.

Several strategies are identified for improving posterior accuracy, each targeting a distinct stage of the inference pipeline. The first one is hyperparameter optimisation: the network architecture and training procedure expose a range of hyperparameters, as well as the parameters for polynomial approximation. The present results are reported for the simplest configuration; a systematic scan over this hyperparameter space, potentially guided by Bayesian optimisation, may reveal configurations that better capture the narrow chirp mass posterior without sacrificing performance on other parameters. A comprehensive comparison of network configurations will be the subject of a separate analysis. The other way is importance sampling, which uses the neural posterior as a proposal distribution and reweights samples to correct for any discrepancy with the exact Bayesian posterior \citep{Dax2023}. This procedure provides a rigorous convergence guarantee: provided the neural posterior has sufficient support overlap with the true posterior, the reweighted samples are guaranteed to converge to exact Bayesian inference in the limit of large sample counts. Although importance sampling reduces the net throughput relative to direct \ac{NPE}, it offers a principled mechanism for correcting systematic biases introduced by the finite capacity of the current network.

Two further limitations deserve discussion. As the \ac{PSD} forms the core of the summary data, the network is expected to exhibit significant bias when the true noise curve deviates substantially from that assumed during training. This sensitivity is an inherent consequence of compressing the data into \ac{PSD}-dependent summary statistics; however, it can be mitigated by training on a distribution of \acp{PSD} drawn from realistic noise realisations, which encourages the network to learn a more robust, noise-marginalised posterior. In addition, the restricted chirp mass prior range can be addressed by adopting the prior-conditioned training scheme of \citet{Dax2025}, in which the network is trained simultaneously on multiple restricted priors tiling the full astrophysical chirp mass range, each paired with an independent heterodyning of the gravitational-wave strain with respect to the representative chirp mass of that prior region.

The inference speed offered by \ac{NPE} fundamentally changes the operational context, particularly for the next-generation detectors. Once trained, the network produces posterior samples in a second per event, enabling real-time parameter estimation for low-latency multi-messenger follow-up—a capability that conventional samplers cannot match within current time constraints. As the sensitivity of ground-based detectors continues to improve, and as the expected detection rate increases accordingly, this speed advantage will become increasingly critical. As detection rates grow with improving detector sensitivity, \ac{NPE}-based inference stands as an essential tool for the emerging high-rate, low-latency era of \ac{GW} astronomy.

\section*{Acknowledgements}

Authors thank Sama Al-Shammari and Alexandre G{\"o}ttel for developing the foundation of the code developed in this study.
M. I. acknowledges support from the Royal Society Award ICA\textbackslash R1\textbackslash 231114. 
S. M. acknowledges support from JSPS Grant-in-Aid for Transformative Research Areas (A) No.~23H04891 and No.~23H04893. V. R. acknowledges support from the UK Science and Technology Facilities Council grant UKRI2489 and the Leverhulme Trust Fellowship IF-2024-038.
The authors are grateful for computational resources provided by the LIGO Lab and supported by National Science Foundation Grants PHY-0757058 and PHY-0823459.


\section*{Data Availability}
The data that support the findings of this study are available on request from the corresponding author.


\bibliographystyle{mnras}
\bibliography{bibs}

@ARTICLE{ZackayRelBin,
       author = {{Zackay}, Barak and {Dai}, Liang and {Venumadhav}, Tejaswi},
        title = "{Relative Binning and Fast Likelihood Evaluation for Gravitational Wave Parameter Estimation}",
      journal = {arXiv e-prints},
         year = 2018,
        month = jun,
          eid = {arXiv:1806.08792},
        pages = {arXiv:1806.08792},
          doi = {10.48550/arXiv.1806.08792},
archivePrefix = {arXiv},
       eprint = {1806.08792},
 primaryClass = {astro-ph.IM},
       adsurl = {https://ui.adsabs.harvard.edu/abs/2018arXiv180608792Z}
}

@article{GW150914,
    author = "Abbott, B. P. and others",
    collaboration = "LIGO Scientific, Virgo",
    title = "{Observation of Gravitational Waves from a Binary Black Hole Merger}",
    eprint = "1602.03837",
    archivePrefix = "arXiv",
    primaryClass = "gr-qc",
    reportNumber = "LIGO-P150914",
    doi = "10.1103/PhysRevLett.116.061102",
    journal = "Phys. Rev. Lett.",
    volume = "116",
    number = "6",
    pages = "061102",
    year = "2016"
}

@article{LIGO,
    author = "Aasi, J. and others",
    collaboration = "LIGO Scientific",
    title = "{Advanced LIGO}",
    eprint = "1411.4547",
    archivePrefix = "arXiv",
    primaryClass = "gr-qc",
    doi = "10.1088/0264-9381/32/7/074001",
    journal = "Class. Quant. Grav.",
    volume = "32",
    pages = "074001",
    year = "2015"
}

@article{VIRGO,
    author = "Acernese, F. and others",
    collaboration = "VIRGO",
    title = "{Advanced Virgo: a second-generation interferometric gravitational wave detector}",
    eprint = "1408.3978",
    archivePrefix = "arXiv",
    primaryClass = "gr-qc",
    doi = "10.1088/0264-9381/32/2/024001",
    journal = "Class. Quant. Grav.",
    volume = "32",
    number = "2",
    pages = "024001",
    year = "2015"
}

@article{KAGRA,
    author = {Akutsu, T. and others},
    title = "{Overview of KAGRA: Detector design and construction history}",
    journal = {Progress of Theoretical and Experimental Physics},
    volume = {2021},
    number = {5},
    pages = {05A101},
    year = {2020},
    month = {08},
    issn = {2050-3911},
    doi = {10.1093/ptep/ptaa125},
    url = {https://doi.org/10.1093/ptep/ptaa125},
    eprint = {https://academic.oup.com/ptep/article-pdf/2021/5/05A101/37974994/ptaa125.pdf},
}

@article{LVK,
    author = "Abbott, B. P. and others",
    collaboration = "KAGRA, LIGO Scientific, Virgo",
    title = "{Prospects for observing and localizing gravitational-wave transients with Advanced LIGO, Advanced Virgo and KAGRA}",
    eprint = "1304.0670",
    archivePrefix = "arXiv",
    primaryClass = "gr-qc",
    reportNumber = "LIGO-P1200087, VIR-0288A-12, JGW-P1808427",
    doi = "10.1007/s41114-020-00026-9",
    journal = "Living Rev. Rel.",
    volume = "19",
    pages = "1",
    year = "2016"
}

@ARTICLE{GWTC-1,
       author = {{Abbott}, B.~P. and others},
collaboration = {{LIGO Scientific Collaboration} and {Virgo Collaboration}},
        title = "{GWTC-1: A Gravitational-Wave Transient Catalog of Compact Binary Mergers Observed by LIGO and Virgo during the First and Second Observing Runs}",
      journal = {Physical Review X},
         year = 2019,
        month = jul,
       volume = {9},
       number = {3},
          eid = {031040},
        pages = {031040},
          doi = {10.1103/PhysRevX.9.031040},
archivePrefix = {arXiv},
       eprint = {1811.12907},
 primaryClass = {astro-ph.HE},
       adsurl = {https://ui.adsabs.harvard.edu/abs/2019PhRvX...9c1040A}
}

@ARTICLE{GWTC-2,
       author = {{Abbott}, R. and others},
collaboration = {{LIGO Scientific Collaboration}, Virgo Collaboration, and {KAGRA Collaboration}},
        title = "{GWTC-2: Compact Binary Coalescences Observed by LIGO and Virgo during the First Half of the Third Observing Run}",
      journal = {Physical Review X},
         year = 2021,
        month = apr,
       volume = {11},
       number = {2},
          eid = {021053},
        pages = {021053},
          doi = {10.1103/PhysRevX.11.021053},
archivePrefix = {arXiv},
       eprint = {2010.14527},
 primaryClass = {gr-qc},
       adsurl = {https://ui.adsabs.harvard.edu/abs/2021PhRvX..11b1053A}
}

@ARTICLE{GWTC-2.1,
       author = {{Abbott}, R. and others},
collaboration = {{LIGO Scientific Collaboration}, Virgo Collaboration, and {KAGRA Collaboration}},
        title = "{GWTC-2.1: Deep extended catalog of compact binary coalescences observed by LIGO and Virgo during the first half of the third observing run}",
      journal = {\prd},
         year = 2024,
        month = jan,
       volume = {109},
       number = {2},
          eid = {022001},
        pages = {022001},
          doi = {10.1103/PhysRevD.109.022001},
archivePrefix = {arXiv},
       eprint = {2108.01045},
 primaryClass = {gr-qc},
       adsurl = {https://ui.adsabs.harvard.edu/abs/2024PhRvD.109b2001A}
}

@ARTICLE{GWTC-3,
       author = {{Abbott}, R. and others},
collaboration = {{LIGO Scientific Collaboration}, Virgo Collaboration, and {KAGRA Collaboration}},
        title = "{GWTC-3: Compact Binary Coalescences Observed by LIGO and Virgo during the Second Part of the Third Observing Run}",
      journal = {Physical Review X},
         year = 2023,
        month = oct,
       volume = {13},
       number = {4},
          eid = {041039},
        pages = {041039},
          doi = {10.1103/PhysRevX.13.041039},
archivePrefix = {arXiv},
       eprint = {2111.03606},
 primaryClass = {gr-qc},
       adsurl = {https://ui.adsabs.harvard.edu/abs/2023PhRvX..13d1039A}
}

@article{GWTC-4,
    author = "Abac, A. G. and others",
    collaboration = "LIGO Scientific, VIRGO, KAGRA",
    title = "{GWTC-4.0: Updating the Gravitational-Wave Transient Catalog with Observations from the First Part of the Fourth LIGO-Virgo-KAGRA Observing Run}",
    eprint = "2508.18082",
    archivePrefix = "arXiv",
    primaryClass = "gr-qc",
    reportNumber = "LIGO-P2400386",
    month = "8",
    journal = "arXiv preprint",
    year = "2025"
}

@ARTICLE{TGR-1,
       author = {{Abbott}, B.~P. and others},
        title = "{Tests of general relativity with the binary black hole signals from the LIGO-Virgo catalog GWTC-1}",
      journal = {\prd},
         year = 2019,
        month = nov,
       volume = {100},
       number = {10},
          eid = {104036},
        pages = {104036},
          doi = {10.1103/PhysRevD.100.104036},
archivePrefix = {arXiv},
       eprint = {1903.04467},
 primaryClass = {gr-qc},
       adsurl = {https://ui.adsabs.harvard.edu/abs/2019PhRvD.100j4036A}
}

@ARTICLE{TGR-2,
       author = {{Abbott}, R. and others},
        title = "{Tests of general relativity with binary black holes from the second LIGO-Virgo gravitational-wave transient catalog}",
      journal = {\prd},
         year = 2021,
        month = jun,
       volume = {103},
       number = {12},
          eid = {122002},
        pages = {122002},
          doi = {10.1103/PhysRevD.103.122002},
archivePrefix = {arXiv},
       eprint = {2010.14529},
 primaryClass = {gr-qc},
       adsurl = {https://ui.adsabs.harvard.edu/abs/2021PhRvD.103l2002A}
}

@ARTICLE{TGR-3,
       author = {{Abbott}, R. and others},
        title = "{Tests of general relativity with GWTC-3}",
      journal = {\prd},
         year = 2025,
        month = oct,
       volume = {112},
       number = {8},
          eid = {084080},
        pages = {084080},
          doi = {10.1103/PhysRevD.112.084080},
archivePrefix = {arXiv},
       eprint = {2112.06861},
 primaryClass = {gr-qc},
       adsurl = {https://ui.adsabs.harvard.edu/abs/2025PhRvD.112h4080A}
}

@ARTICLE{RP-1,
       author = {{Abbott}, B.~P. and others},
collaboration = {{LIGO Scientific Collaboration} and {Virgo Collaboration}},
        title = "{Binary Black Hole Population Properties Inferred from the First and Second Observing Runs of Advanced LIGO and Advanced Virgo}",
      journal = {\apjl},
         year = 2019,
        month = sep,
       volume = {882},
       number = {2},
          eid = {L24},
        pages = {L24},
          doi = {10.3847/2041-8213/ab3800},
archivePrefix = {arXiv},
       eprint = {1811.12940},
 primaryClass = {astro-ph.HE},
       adsurl = {https://ui.adsabs.harvard.edu/abs/2019ApJ...882L..24A}
}

@ARTICLE{RP-2,
       author = {{Abbott}, R. and others},
collaboration = {{LIGO Scientific Collaboration} and {Virgo Collaboration}},
        title = "{Population Properties of Compact Objects from the Second LIGO-Virgo Gravitational-Wave Transient Catalog}",
      journal = {\apjl},
         year = 2021,
        month = may,
       volume = {913},
       number = {1},
          eid = {L7},
        pages = {L7},
          doi = {10.3847/2041-8213/abe949},
archivePrefix = {arXiv},
       eprint = {2010.14533},
 primaryClass = {astro-ph.HE},
       adsurl = {https://ui.adsabs.harvard.edu/abs/2021ApJ...913L...7A}
}

@ARTICLE{RP-3,
       author = {{Abbott}, R. and others},
collaboration = {{LIGO Scientific Collaboration} and {VIRGO Collaboration} and {KAGRA Collaboration}},
        title = "{Population of Merging Compact Binaries Inferred Using Gravitational Waves through GWTC-3}",
      journal = {Physical Review X},
         year = 2023,
        month = jan,
       volume = {13},
       number = {1},
          eid = {011048},
        pages = {011048},
          doi = {10.1103/PhysRevX.13.011048},
archivePrefix = {arXiv},
       eprint = {2111.03634},
 primaryClass = {astro-ph.HE},
       adsurl = {https://ui.adsabs.harvard.edu/abs/2023PhRvX..13a1048A}
}

@article{RP-4,
    author = "Abac, A. G. and others",
    collaboration = "LIGO Scientific, VIRGO, KAGRA",
    title = "{GWTC-4.0: Population Properties of Merging Compact Binaries}",
    eprint = "2508.18083",
    archivePrefix = "arXiv",
    primaryClass = "astro-ph.HE",
    reportNumber = "LIGO-P2400004",
    month = "8",
    journal = "arXiv preprint",
    year = "2025"
}

@ARTICLE{GW170817,
       author = {{Abbott}, B.~P. and others},
        title = "{GW170817: Observation of Gravitational Waves from a Binary Neutron Star Inspiral}",
      journal = {\prl},
         year = 2017,
        month = oct,
       volume = {119},
       number = {16},
          eid = {161101},
        pages = {161101},
          doi = {10.1103/PhysRevLett.119.161101},
archivePrefix = {arXiv},
       eprint = {1710.05832},
 primaryClass = {gr-qc},
       adsurl = {https://ui.adsabs.harvard.edu/abs/2017PhRvL.119p1101A}
}

@ARTICLE{GW170817MM,
       author = {{Abbott}, B.~P. and others},
        title = "{Multi-messenger Observations of a Binary Neutron Star Merger}",
      journal = {\apjl},
         year = 2017,
        month = oct,
       volume = {848},
       number = {2},
          eid = {L12},
        pages = {L12},
          doi = {10.3847/2041-8213/aa91c9},
archivePrefix = {arXiv},
       eprint = {1710.05833},
 primaryClass = {astro-ph.HE},
       adsurl = {https://ui.adsabs.harvard.edu/abs/2017ApJ...848L..12A}
}

@ARTICLE{GW170817:r-process,
       author = {{Pian}, E. and others},
        title = "{Spectroscopic identification of r-process nucleosynthesis in a double neutron-star merger}",
      journal = {\nat},
         year = 2017,
        month = nov,
       volume = {551},
       number = {7678},
        pages = {67-70},
          doi = {10.1038/nature24298},
archivePrefix = {arXiv},
       eprint = {1710.05858},
 primaryClass = {astro-ph.HE},
       adsurl = {https://ui.adsabs.harvard.edu/abs/2017Natur.551...67P}
}

@ARTICLE{GW170817:siren,
       author = {{Abbott}, B.~P. and others},
        title = "{A gravitational-wave standard siren measurement of the Hubble constant}",
      journal = {\nat},
         year = 2017,
        month = nov,
       volume = {551},
       number = {7678},
        pages = {85-88},
          doi = {10.1038/nature24471},
archivePrefix = {arXiv},
       eprint = {1710.05835},
 primaryClass = {astro-ph.CO},
       adsurl = {https://ui.adsabs.harvard.edu/abs/2017Natur.551...85A}
}

@ARTICLE{GW170817:EoS,
       author = {{Abbott}, B.~P. and others},
        title = "{GW170817: Measurements of Neutron Star Radii and Equation of State}",
      journal = {\prl},
         year = 2018,
        month = oct,
       volume = {121},
       number = {16},
          eid = {161101},
        pages = {161101},
          doi = {10.1103/PhysRevLett.121.161101},
archivePrefix = {arXiv},
       eprint = {1805.11581},
 primaryClass = {gr-qc},
       adsurl = {https://ui.adsabs.harvard.edu/abs/2018PhRvL.121p1101A}
}

@INPROCEEDINGS{NestedSampling,
       author = {{Skilling}, John},
        title = "{Nested Sampling}",
    booktitle = {Bayesian Inference and Maximum Entropy Methods in Science and Engineering: 24th International Workshop on Bayesian Inference and Maximum Entropy Methods in Science and Engineering},
         year = 2004,
       editor = {{Fischer}, Rainer and {Preuss}, Roland and {Toussaint}, Udo Von},
       series = {American Institute of Physics Conference Series},
       volume = {735},
        month = nov,
    publisher = {AIP},
        pages = {395-405},
          doi = {10.1063/1.1835238},
       adsurl = {https://ui.adsabs.harvard.edu/abs/2004AIPC..735..395S}
}

@ARTICLE{LALinference,
       author = {{Veitch}, J. and {Raymond}, V. and {Farr}, B. and {Farr}, W. and {Graff}, P. and {Vitale}, S. and {Aylott}, B. and {Blackburn}, K. and {Christensen}, N. and {Coughlin}, M. and {Del Pozzo}, W. and {Feroz}, F. and {Gair}, J. and {Haster}, C.-J. and {Kalogera}, V. and {Littenberg}, T. and {Mandel}, I. and {O'Shaughnessy}, R. and {Pitkin}, M. and {Rodriguez}, C. and {R{\"o}ver}, C. and {Sidery}, T. and {Smith}, R. and {Van Der Sluys}, M. and {Vecchio}, A. and {Vousden}, W. and {Wade}, L.},
        title = "{Parameter estimation for compact binaries with ground-based gravitational-wave observations using the LALInference software library}",
      journal = {\prd},
         year = 2015,
        month = feb,
       volume = {91},
       number = {4},
          eid = {042003},
        pages = {042003},
          doi = {10.1103/PhysRevD.91.042003},
archivePrefix = {arXiv},
       eprint = {1409.7215},
 primaryClass = {gr-qc},
       adsurl = {https://ui.adsabs.harvard.edu/abs/2015PhRvD..91d2003V}
}

@ARTICLE{Bilby,
       author = {{Ashton}, Gregory and {H{\"u}bner}, Moritz and {Lasky}, Paul D. and {Talbot}, Colm and {Ackley}, Kendall and {Biscoveanu}, Sylvia and {Chu}, Qi and {Divakarla}, Atul and {Easter}, Paul J. and {Goncharov}, Boris and {Hernandez Vivanco}, Francisco and {Harms}, Jan and {Lower}, Marcus E. and {Meadors}, Grant D. and {Melchor}, Denyz and {Payne}, Ethan and {Pitkin}, Matthew D. and {Powell}, Jade and {Sarin}, Nikhil and {Smith}, Rory J.~E. and {Thrane}, Eric},
        title = "{BILBY: A User-friendly Bayesian Inference Library for Gravitational-wave Astronomy}",
      journal = {\apjs},
         year = 2019,
        month = apr,
       volume = {241},
       number = {2},
          eid = {27},
        pages = {27},
          doi = {10.3847/1538-4365/ab06fc},
archivePrefix = {arXiv},
       eprint = {1811.02042},
 primaryClass = {astro-ph.IM},
       adsurl = {https://ui.adsabs.harvard.edu/abs/2019ApJS..241...27A}
}

@ARTICLE{CranmerSBI,
       author = {{Cranmer}, Kyle and {Brehmer}, Johann and {Louppe}, Gilles},
        title = "{The frontier of simulation-based inference}",
      journal = {Proceedings of the National Academy of Science},
         year = 2020,
        month = dec,
       volume = {117},
       number = {48},
        pages = {30055-30062},
          doi = {10.1073/pnas.1912789117},
archivePrefix = {arXiv},
       eprint = {1911.01429},
 primaryClass = {stat.ML},
       adsurl = {https://ui.adsabs.harvard.edu/abs/2020PNAS..11730055C}
}

@ARTICLE{Green2020nov,
       author = {{Green}, Stephen R. and {Simpson}, Christine and {Gair}, Jonathan},
        title = "{Gravitational-wave parameter estimation with autoregressive neural network flows}",
      journal = {\prd},
         year = 2020,
        month = nov,
       volume = {102},
       number = {10},
          eid = {104057},
        pages = {104057},
          doi = {10.1103/PhysRevD.102.104057},
archivePrefix = {arXiv},
       eprint = {2002.07656},
 primaryClass = {astro-ph.IM},
       adsurl = {https://ui.adsabs.harvard.edu/abs/2020PhRvD.102j4057G}
}

@ARTICLE{Delaunoy2020,
       author = {{Delaunoy}, Arnaud and {Wehenkel}, Antoine and {Hinderer}, Tanja and {Nissanke}, Samaya and {Weniger}, Christoph and {Williamson}, Andrew R. and {Louppe}, Gilles},
        title = "{Lightning-Fast Gravitational Wave Parameter Inference through Neural Amortization}",
      journal = {arXiv e-prints},
         year = 2020,
        month = oct,
          eid = {arXiv:2010.12931},
        pages = {arXiv:2010.12931},
          doi = {10.48550/arXiv.2010.12931},
archivePrefix = {arXiv},
       eprint = {2010.12931},
 primaryClass = {astro-ph.IM},
       adsurl = {https://ui.adsabs.harvard.edu/abs/2020arXiv201012931D}
}

@ARTICLE{Green2020aug,
       author = {{Green}, Stephen R. and {Gair}, Jonathan},
        title = "{Complete parameter inference for GW150914 using deep learning}",
      journal = {arXiv e-prints},
         year = 2020,
        month = aug,
          eid = {arXiv:2008.03312},
        pages = {arXiv:2008.03312},
          doi = {10.48550/arXiv.2008.03312},
archivePrefix = {arXiv},
       eprint = {2008.03312},
 primaryClass = {astro-ph.IM},
       adsurl = {https://ui.adsabs.harvard.edu/abs/2020arXiv200803312G}
}

@ARTICLE{Dax2021,
       author = {{Dax}, Maximilian and {Green}, Stephen R. and {Gair}, Jonathan and {Macke}, Jakob H. and {Buonanno}, Alessandra and {Sch{\"o}lkopf}, Bernhard},
        title = "{Real-Time Gravitational Wave Science with Neural Posterior Estimation}",
      journal = {\prl},
         year = 2021,
        month = dec,
       volume = {127},
       number = {24},
          eid = {241103},
        pages = {241103},
          doi = {10.1103/PhysRevLett.127.241103},
archivePrefix = {arXiv},
       eprint = {2106.12594},
 primaryClass = {gr-qc},
       adsurl = {https://ui.adsabs.harvard.edu/abs/2021PhRvL.127x1103D}
}

@ARTICLE{Dax2023,
       author = {{Dax}, Maximilian and {Green}, Stephen R. and {Gair}, Jonathan and {P{\"u}rrer}, Michael and {Wildberger}, Jonas and {Macke}, Jakob H. and {Buonanno}, Alessandra and {Sch{\"o}lkopf}, Bernhard},
        title = "{Neural Importance Sampling for Rapid and Reliable Gravitational-Wave Inference}",
      journal = {\prl},
         year = 2023,
        month = apr,
       volume = {130},
       number = {17},
          eid = {171403},
        pages = {171403},
          doi = {10.1103/PhysRevLett.130.171403},
archivePrefix = {arXiv},
       eprint = {2210.05686},
 primaryClass = {gr-qc},
       adsurl = {https://ui.adsabs.harvard.edu/abs/2023PhRvL.130q1403D}
}

@ARTICLE{Bhardwaj2023,
       author = {{Bhardwaj}, Uddipta and {Alvey}, James and {Miller}, Benjamin Kurt and {Nissanke}, Samaya and {Weniger}, Christoph},
        title = "{Sequential simulation-based inference for gravitational wave signals}",
      journal = {\prd},
         year = 2023,
        month = aug,
       volume = {108},
       number = {4},
          eid = {042004},
        pages = {042004},
          doi = {10.1103/PhysRevD.108.042004},
archivePrefix = {arXiv},
       eprint = {2304.02035},
 primaryClass = {gr-qc},
       adsurl = {https://ui.adsabs.harvard.edu/abs/2023PhRvD.108d2004B}
}

@ARTICLE{Alvey2023,
       author = {{Alvey}, James and {Bhardwaj}, Uddipta and {Nissanke}, Samaya and {Weniger}, Christoph},
        title = "{What to do when things get crowded? Scalable joint analysis of overlapping gravitational wave signals}",
      journal = {arXiv e-prints},
         year = 2023,
        month = aug,
          eid = {arXiv:2308.06318},
        pages = {arXiv:2308.06318},
          doi = {10.48550/arXiv.2308.06318},
archivePrefix = {arXiv},
       eprint = {2308.06318},
 primaryClass = {gr-qc},
       adsurl = {https://ui.adsabs.harvard.edu/abs/2023arXiv230806318A}
}

@ARTICLE{Dax2025,
       author = {{Dax}, Maximilian and {Green}, Stephen R. and {Gair}, Jonathan and {Gupte}, Nihar and {P{\"u}rrer}, Michael and {Raymond}, Vivien and {Wildberger}, Jonas and {Macke}, Jakob H. and {Buonanno}, Alessandra and {Sch{\"o}lkopf}, Bernhard},
        title = "{Real-time inference for binary neutron star mergers using machine learning}",
      journal = {\nat},
         year = 2025,
        month = mar,
       volume = {639},
       number = {8053},
        pages = {49-53},
          doi = {10.1038/s41586-025-08593-z},
archivePrefix = {arXiv},
       eprint = {2407.09602},
 primaryClass = {gr-qc},
       adsurl = {https://ui.adsabs.harvard.edu/abs/2025Natur.639...49D}
}

@ARTICLE{Hu2025,
       author = {{Hu}, Qian and {Irwin}, Jessica and {Sun}, Qi and {Messenger}, Christopher and {Suleiman}, Lami and {Heng}, Ik Siong and {Veitch}, John},
        title = "{Decoding Long-duration Gravitational Waves from Binary Neutron Stars with Machine Learning: Parameter Estimation and Equations of State}",
      journal = {\apjl},
         year = 2025,
        month = jul,
       volume = {987},
       number = {1},
          eid = {L17},
        pages = {L17},
          doi = {10.3847/2041-8213/ade42f},
archivePrefix = {arXiv},
       eprint = {2412.03454},
 primaryClass = {gr-qc},
       adsurl = {https://ui.adsabs.harvard.edu/abs/2025ApJ...987L..17H}
}

@ARTICLE{GlitchJade2018,
       author = {{Powell}, Jade},
        title = "{Parameter estimation and model selection of gravitational wave signals contaminated by transient detector noise glitches}",
      journal = {Classical and Quantum Gravity},
         year = 2018,
        month = aug,
       volume = {35},
       number = {15},
          eid = {155017},
        pages = {155017},
          doi = {10.1088/1361-6382/aacf18},
archivePrefix = {arXiv},
       eprint = {1803.11346},
 primaryClass = {astro-ph.IM},
       adsurl = {https://ui.adsabs.harvard.edu/abs/2018CQGra..35o5017P}
}

@ARTICLE{GlitchPayne2022,
       author = {{Payne}, Ethan and {Hourihane}, Sophie and {Golomb}, Jacob and {Udall}, Rhiannon and {Davis}, Derek and {Chatziioannou}, Katerina},
        title = "{Curious case of GW200129: Interplay between spin-precession inference and data-quality issues}",
      journal = {\prd},
         year = 2022,
        month = nov,
       volume = {106},
       number = {10},
          eid = {104017},
        pages = {104017},
          doi = {10.1103/PhysRevD.106.104017},
archivePrefix = {arXiv},
       eprint = {2206.11932},
 primaryClass = {gr-qc},
       adsurl = {https://ui.adsabs.harvard.edu/abs/2022PhRvD.106j4017P}
}

@ARTICLE{GlitchUdall2025,
       author = {{Udall}, Rhiannon and {Hourihane}, Sophie and {Miller}, Simona and {Davis}, Derek and {Chatziioannou}, Katerina and {Isi}, Max and {Deshong}, Howard},
        title = "{Antialigned spin of GW191109: Glitch mitigation and its implications}",
      journal = {\prd},
         year = 2025,
        month = jan,
       volume = {111},
       number = {2},
          eid = {024046},
        pages = {024046},
          doi = {10.1103/PhysRevD.111.024046},
archivePrefix = {arXiv},
       eprint = {2409.03912},
 primaryClass = {gr-qc},
       adsurl = {https://ui.adsabs.harvard.edu/abs/2025PhRvD.111b4046U}
}

@ARTICLE{Heterodyne,
       author = {{Cornish}, Neil J.},
        title = "{Fast Fisher Matrices and Lazy Likelihoods}",
      journal = {arXiv e-prints},
         year = 2010,
        month = jul,
          eid = {arXiv:1007.4820},
        pages = {arXiv:1007.4820},
          doi = {10.48550/arXiv.1007.4820},
archivePrefix = {arXiv},
       eprint = {1007.4820},
 primaryClass = {gr-qc},
       adsurl = {https://ui.adsabs.harvard.edu/abs/2010arXiv1007.4820C}
}

@ARTICLE{Multiband1,
       author = {{Vinciguerra}, Serena and {Veitch}, John and {Mandel}, Ilya},
        title = "{Accelerating gravitational wave parameter estimation with multi-band template interpolation}",
      journal = {Classical and Quantum Gravity},
         year = 2017,
        month = jun,
       volume = {34},
       number = {11},
          eid = {115006},
        pages = {115006},
          doi = {10.1088/1361-6382/aa6d44},
archivePrefix = {arXiv},
       eprint = {1703.02062},
 primaryClass = {gr-qc},
       adsurl = {https://ui.adsabs.harvard.edu/abs/2017CQGra..34k5006V}
}

@ARTICLE{Multiband2,
       author = {{Morisaki}, Soichiro},
        title = "{Accelerating parameter estimation of gravitational waves from compact binary coalescence using adaptive frequency resolutions}",
      journal = {\prd},
         year = 2021,
        month = aug,
       volume = {104},
       number = {4},
          eid = {044062},
        pages = {044062},
          doi = {10.1103/PhysRevD.104.044062},
archivePrefix = {arXiv},
       eprint = {2104.07813},
 primaryClass = {gr-qc},
       adsurl = {https://ui.adsabs.harvard.edu/abs/2021PhRvD.104d4062M}
}

@ARTICLE{Likelihood1,
       author = {{van der Sluys}, Marc and {Raymond}, Vivien and {Mandel}, Ilya and {R{\"o}ver}, Christian and {Christensen}, Nelson and {Kalogera}, Vicky and {Meyer}, Renate and {Vecchio}, Alberto},
        title = "{Parameter estimation of spinning binary inspirals using Markov chain Monte Carlo}",
      journal = {Classical and Quantum Gravity},
         year = 2008,
        month = sep,
       volume = {25},
       number = {18},
          eid = {184011},
        pages = {184011},
          doi = {10.1088/0264-9381/25/18/184011},
archivePrefix = {arXiv},
       eprint = {0805.1689},
 primaryClass = {gr-qc},
       adsurl = {https://ui.adsabs.harvard.edu/abs/2008CQGra..25r4011V}
}

@ARTICLE{Likelihood2,
       author = {{van der Sluys}, M.~V. and {R{\"o}ver}, C. and {Stroeer}, A. and {Raymond}, V. and {Mandel}, I. and {Christensen}, N. and {Kalogera}, V. and {Meyer}, R. and {Vecchio}, A.},
        title = "{Gravitational-Wave Astronomy with Inspiral Signals of Spinning Compact-Object Binaries}",
      journal = {\apjl},
         year = 2008,
        month = dec,
       volume = {688},
       number = {2},
        pages = {L61},
          doi = {10.1086/595279},
archivePrefix = {arXiv},
       eprint = {0710.1897},
 primaryClass = {astro-ph},
       adsurl = {https://ui.adsabs.harvard.edu/abs/2008ApJ...688L..61V}
}

@ARTICLE{Likelihood3,
       author = {{Veitch}, J. and {Vecchio}, A.},
        title = "{Bayesian approach to the follow-up of candidate gravitational wave signals}",
      journal = {\prd},
         year = 2008,
        month = jul,
       volume = {78},
       number = {2},
          eid = {022001},
        pages = {022001},
          doi = {10.1103/PhysRevD.78.022001},
archivePrefix = {arXiv},
       eprint = {0801.4313},
 primaryClass = {gr-qc},
       adsurl = {https://ui.adsabs.harvard.edu/abs/2008PhRvD..78b2001V}
}

@article{SBIpackage,
  doi = {10.21105/joss.07754},
  url = {https://doi.org/10.21105/joss.07754},
  year = {2025},
  publisher = {The Open Journal},
  volume = {10},
  number = {108},
  pages = {7754},
  author = {Jan Boelts and Michael Deistler and Manuel Gloeckler and Álvaro Tejero-Cantero and Jan-Matthis Lueckmann and Guy Moss and Peter Steinbach and Thomas Moreau and Fabio Muratore and Julia Linhart and Conor Durkan and Julius Vetter and Benjamin Kurt Miller and Maternus Herold and Abolfazl Ziaeemehr and Matthijs Pals and Theo Gruner and Sebastian Bischoff and Nastya Krouglova and Richard Gao and Janne K. Lappalainen and Bálint Mucsányi and Felix Pei and Auguste Schulz and Zinovia Stefanidi and Pedro Rodrigues and Cornelius Schröder and Faried Abu Zaid and Jonas Beck and Jaivardhan Kapoor and David S. Greenberg and Pedro J. Gonçalves and Jakob H. Macke},
  title = {sbi reloaded: a toolkit for simulation-based inference workflows},
  journal = {Journal of Open Source Software}
}

@ARTICLE{bilby_pipe_paper,
       author = {{Romero-Shaw}, I.~M. and {Talbot}, C. and {Biscoveanu}, S. and {D'Emilio}, V. and {Ashton}, G. and {Berry}, C.~P.~L. and {Coughlin}, S. and {Galaudage}, S. and {Hoy}, C. and {H{\"u}bner}, M. and {Phukon}, K.~S. and {Pitkin}, M. and {Rizzo}, M. and {Sarin}, N. and {Smith}, R. and {Stevenson}, S. and {Vajpeyi}, A. and {Ar{\`e}ne}, M. and {Athar}, K. and {Banagiri}, S. and {Bose}, N. and {Carney}, M. and {Chatziioannou}, K. and {Clark}, J.~A. and {Colleoni}, M. and {Cotesta}, R. and {Edelman}, B. and {Estell{\'e}s}, H. and {Garc{\'\i}a-Quir{\'o}s}, C. and {Ghosh}, Abhirup and {Green}, R. and {Haster}, C.-J. and {Husa}, S. and {Keitel}, D. and {Kim}, A.~X. and {Hernandez-Vivanco}, F. and {Maga{\~n}a Hernandez}, I. and {Karathanasis}, C. and {Lasky}, P.~D. and {De Lillo}, N. and {Lower}, M.~E. and {Macleod}, D. and {Mateu-Lucena}, M. and {Miller}, A. and {Millhouse}, M. and {Morisaki}, S. and {Oh}, S.~H. and {Ossokine}, S. and {Payne}, E. and {Powell}, J. and {Pratten}, G. and {P{\"u}rrer}, M. and {Ramos-Buades}, A. and {Raymond}, V. and {Thrane}, E. and {Veitch}, J. and {Williams}, D. and {Williams}, M.~J. and {Xiao}, L.},
        title = "{Bayesian inference for compact binary coalescences with BILBY: validation and application to the first LIGO-Virgo gravitational-wave transient catalogue}",
      journal = {\mnras},
         year = 2020,
        month = dec,
       volume = {499},
       number = {3},
        pages = {3295-3319},
          doi = {10.1093/mnras/staa2850},
archivePrefix = {arXiv},
       eprint = {2006.00714},
 primaryClass = {astro-ph.IM},
       adsurl = {https://ui.adsabs.harvard.edu/abs/2020MNRAS.499.3295R}
}

@ARTICLE{Dynesty,
       author = {{Speagle}, Joshua S.},
        title = "{DYNESTY: a dynamic nested sampling package for estimating Bayesian posteriors and evidences}",
      journal = {\mnras},
         year = 2020,
        month = apr,
       volume = {493},
       number = {3},
        pages = {3132-3158},
          doi = {10.1093/mnras/staa278},
archivePrefix = {arXiv},
       eprint = {1904.02180},
 primaryClass = {astro-ph.IM},
       adsurl = {https://ui.adsabs.harvard.edu/abs/2020MNRAS.493.3132S}
}

@ARTICLE{NSF,
       author = {{Durkan}, Conor and {Bekasov}, Artur and {Murray}, Iain and {Papamakarios}, George},
        title = "{Neural Spline Flows}",
      journal = {arXiv e-prints},
         year = 2019,
        month = jun,
          eid = {arXiv:1906.04032},
        pages = {arXiv:1906.04032},
          doi = {10.48550/arXiv.1906.04032},
archivePrefix = {arXiv},
       eprint = {1906.04032},
 primaryClass = {stat.ML},
       adsurl = {https://ui.adsabs.harvard.edu/abs/2019arXiv190604032D}
}

@ARTICLE{VRpastwork,
       author = {{Raymond}, Vivien and {Al-Shammari}, Sama and {G{\"o}ttel}, Alexandre},
        title = "{Simulation-based inference for gravitational-waves from intermediate-mass binary black holes in real noise}",
      journal = {\mnras},
         year = 2025,
        month = sep,
       volume = {542},
       number = {2},
        pages = {1103-1108},
          doi = {10.1093/mnras/staf1282},
archivePrefix = {arXiv},
       eprint = {2406.03935},
 primaryClass = {gr-qc},
       adsurl = {https://ui.adsabs.harvard.edu/abs/2025MNRAS.542.1103R}
}



\appendix

\section{Figures for all parameters}
\label{sec: Appendix}

\begin{figure*}
	\includegraphics[width=2\columnwidth]{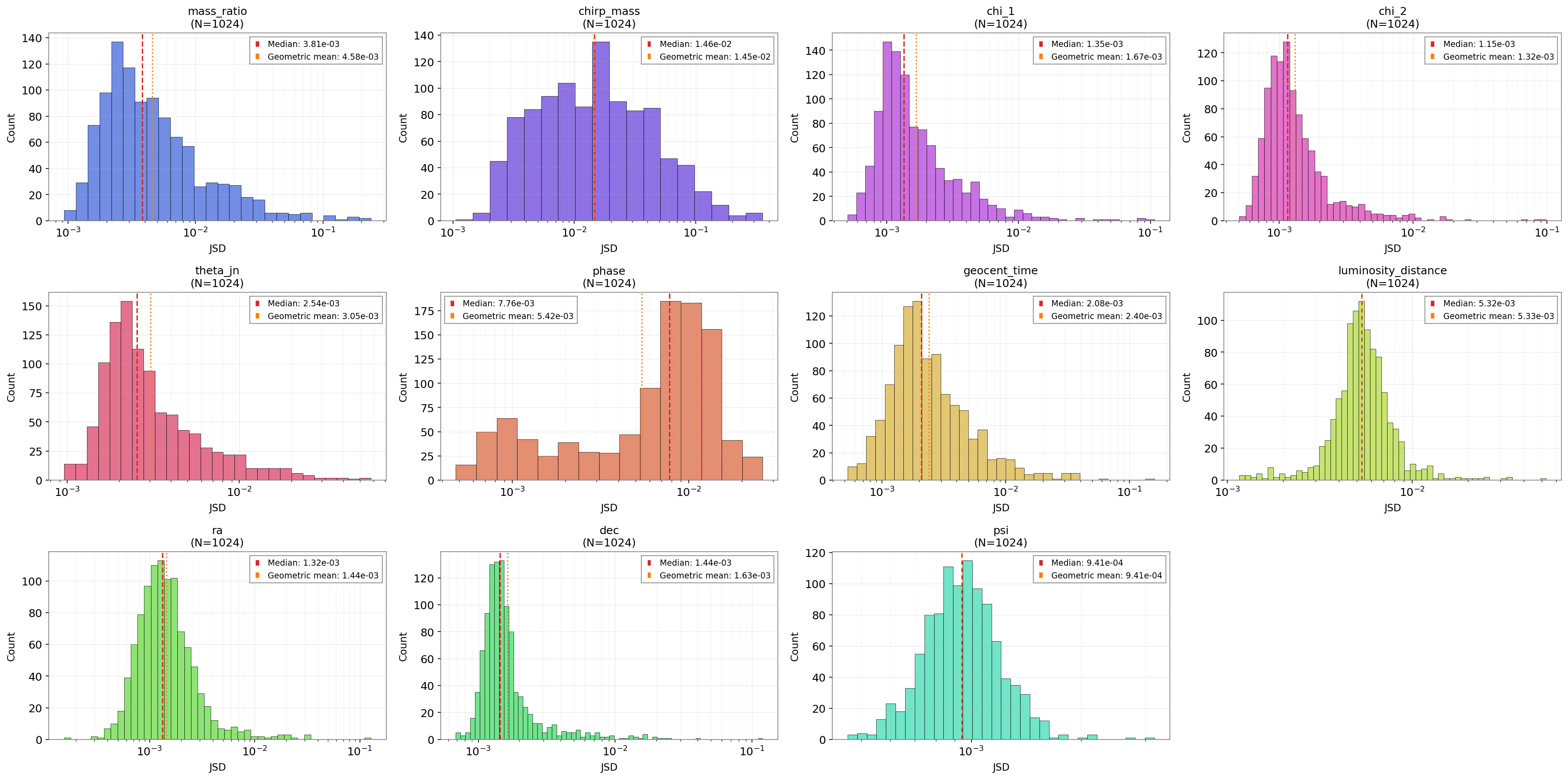}
    \caption{Same as \figref{fig: JSD_histogram}, but for every parameters for the inference.}
    \label{fig: JSD_histogram_full}
\end{figure*}

\figref{fig: JSD_histogram_full} shows the distribution of the \ac{JSD} for all parameters examined in this study. For most parameters, the \ac{JSD} distribution indicates that the SBI network is functioning appropriately. Of particular interest among the newly presented histograms is the phase distribution, which exhibits a slightly bimodal characteristic. This is thought to be because the \ac{SBI} network is sometimes unable to reproduce the wavy characteristics that this parameter occasionally displays in\bilby posterior.

\begin{figure*}
	\includegraphics[width=2\columnwidth]{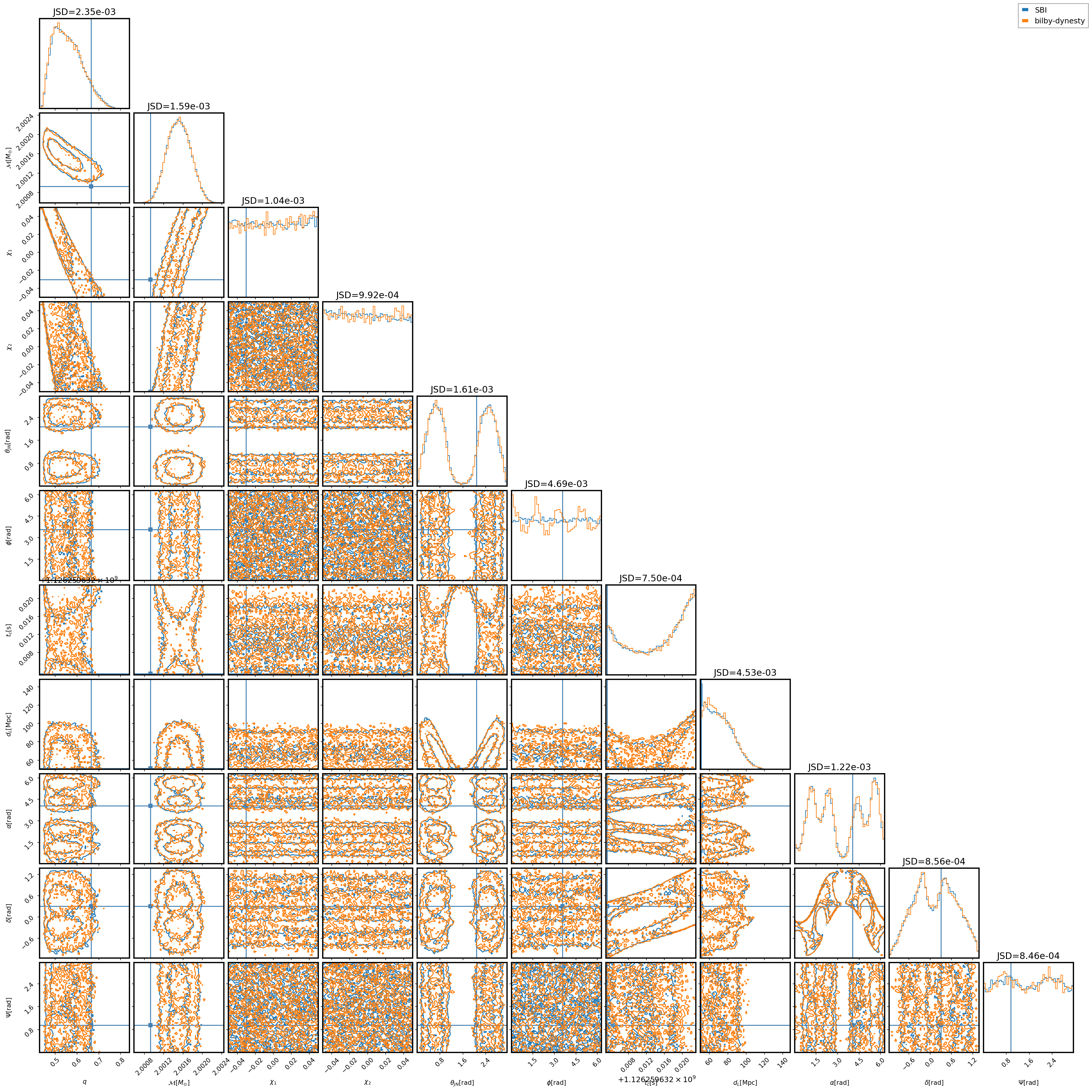}
    \caption{Full corner plot for a BNS injection represented in \figref{fig: Best_chirp_mass}.} 
    \label{fig: Best_chirp_mass_full}
\end{figure*}
\figref{fig: Best_chirp_mass_full} shows the complete corner plot for the same injection as that shown in \figref{fig: Best_chirp_mass}. Again, almost all features shown in the plot agree. The clear exception here is the phase parameter, where the neural posterior fails to reproduce the wavy characteristics in the \bilby posterior. Since we assume a single-detector observation, the coalescence phase is largely unconstrained and less physically informative compared to multi-detector analyses.

\section{Importance sampling with our network}
\label{sec: Appendix_IS}
\begin{figure*}
	\includegraphics[width=2\columnwidth]{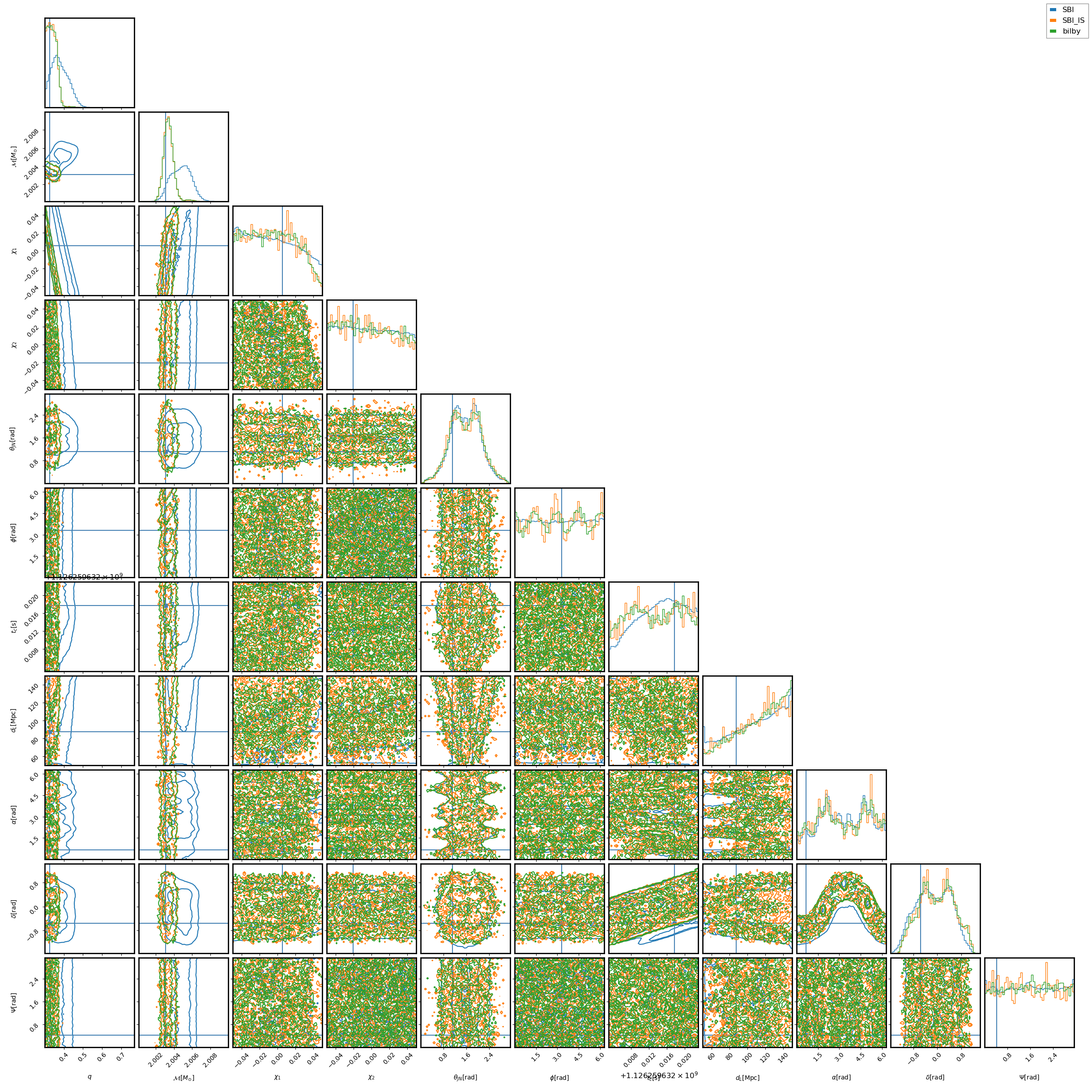}
    \caption{Corner plot for a BNS injection with importance sampling (shown in orange) and raw posterior samples from \ac{NPE} (blue) and \bilby posterior samples (green). For this injection, the JSD between our raw \ac{NPE} samples and the conventional method marks the worst JSD $(0.355)$ obtained in this study.} 
    \label{fig: IS_demonstration}
\end{figure*}
We evaluate the effect of importance sampling using our network, focusing on the injection that produced the worst mass JSD across all tested injections. As shown in \figref{fig: IS_demonstration}, importance sampling substantially improved the posterior accuracy for this injection: the mass JSD decreased from approximately 0.355 bits (raw \ac{NPE} samples) to $1.05\times10^{-2}$ bits. The importance sampling efficiency was approximately 0.2\%, corresponding to an approximately 40-fold reduction in JSD, confirming that importance sampling can effectively correct the residual bias in the network posterior even in the most challenging cases.
This calculation uses $10^6$ raw samples and requires approximately 12 CPU hours of computation time (without multibanding applied to the \textsc{Bilby} likelihood calculation). The $10^6$ raw samples can be obtained within 80 seconds on the LVK computing resource. Although this is faster than a standard \textsc{Bilby} run, the throughput advantage over direct \ac{NPE} is substantially reduced, consistent with the expected trade-off between accuracy and speed discussed in the main text.


\bsp	
\label{lastpage}
\end{document}